\documentclass[preprint,preprintnumbers,amsmath,amssymb,10pt]{revtex4}
\usepackage[colorlinks,linkcolor=red,citecolor=blue]{hyperref}
\usepackage{epsf,epsfig}
\usepackage{bm}
\usepackage{graphicx}
\usepackage{natbib}
\usepackage{color}
\usepackage{mathrsfs}
\usepackage{subfigure}
\usepackage{multirow}
\usepackage{makecell}
\usepackage{booktabs}
\usepackage{doi}

\newcommand{\commentold}[1]{}
\DeclareMathSymbol{:}{\mathpunct}{operators}{"3A}
\begin{document}

%%%%%%%%%%%%%%%%%%%%%%%%%%%%%%%%%%%%%%%%%%%%%%%

\title{Tsallis holographic inflation in $f(R,T)$ gravity: CMB constraints, reheating, and swampland implications}

%%%%%%%%%%%%%%%%%%%%%%%%%%%%%%%%%%%%%%%%%%%%%%%
\author{S. Taghavi}
\thanks{Email: \href{mailto:s.taghavi@uok.ac.ir}{s.taghavi@uok.ac.ir};
	ORCID: \href{https://orcid.org/0009-0005-1121-5562}{0009-0005-1121-5562}}

\author{T. Golanbari}
\thanks{Email: \href{mailto:t.golanbari@uok.ac.ir}{t.golanbari@uok.ac.ir},
	\href{mailto:t.golanbari@gmail.com}{t.golanbari@gmail.com};
	ORCID: \href{https://orcid.org/0000-0003-0973-7335}{0000-0003-0973-7335}(Corresponding author)}

\author{Kh. Saaidi}
\thanks{Email: \href{mailto:ksaaidi@uok.ac.ir}{ksaaidi@uok.ac.ir};
	ORCID: \href{https://orcid.org/0000-0003-0443-8467}{0000-0003-0443-8467}}

\affiliation{Department of Physics, University of Kurdistan, P.O. Box 66177-15175, Sanandaj, Iran}
%%%%%%%%%%%%%%%%%%%%%%%%%%%%%%%%%%%%%%%%%%%%%%%
\date{\today}

%%%%%%%%%%%%%%%%%%%%%%%%%%%%%%%%%%%%%%%%%%%%%%%
%%%%%%%%%%%%%%%%%%%%%%%%%%%%%%%%%%%%%%%%%%%%%%%
%%%%%%%%%%%%%%%%%%%%%%%%%%%%%%%%%%%%%%%%%%%%%%%
%%%%%%%%%%%%%%%%%%%%%%%%%%%%%%%%%%%%%%%%%%%%%%%
% ============ abstract =======================================
%%%%%%%%%%%%%%%%%%%%%%%%%%%%%%%%%%%%%%%%%%%%%%%
%%%%%%%%%%%%%%%%%%%%%%%%%%%%%%%%%%%%%%%%%%%%%%%
%%%%%%%%%%%%%%%%%%%%%%%%%%%%%%%%%%%%%%%%%%%%%%%
%%%%%%%%%%%%%%%%%%%%%%%%%%%%%%%%%%%%%%%%%%%%%%%
\begin{abstract}
		Understanding how early-universe inflation may emerge from generalized holographic energy densities within modified gravity remains one of the intriguing topics in theoretical cosmology, providing strong motivation for the present analysis.
		In this work, we develop a self-consistent inflationary scenario in which the Tsallis holographic dark energy (THDE) density effectively plays the role of the inflaton potential in $f(R,T)$ gravity.
		Using the Granda–Oliveros infrared cutoff, we derive the corresponding slow-roll relations and identify a broad region of the parameter space $(\alpha,\beta,\delta,\lambda)$ that yields scalar perturbation observables consistent with the combined ACT DR6 (P-ACT-LB) likelihoods. \
		
		By exploiting the parametric dependence of the THDE density on the Hubble rate, we reconstruct the inflaton potential $V(\phi)$ over the observable window and show that both the field excursion and the potential gradient normalized by the potential value, quantified by $|V'|/(V M_{p})$, are predominantly controlled by the matter–geometry coupling $\lambda$.
		In particular, we demonstrate that $\lambda \gtrsim \mathcal{O}(10^{2})$ simultaneously suppresses the field excursion below the Planck scale and ensures $|V'|/(V M_{p}) \ge 1$, which satisfies both the distance conjecture and the refined de Sitter swampland bound. \
		
		We also analyze the reheating stage.
		In addition to the primordial nucleosynthesis requirement $T_{\rm BBN} \simeq 4$ MeV, which provides a lower limit on the reheating temperature, the observational bound $\Delta N_{\rm eff} \le 0.17$ imposes an additional constraint arising from primordial gravitational waves (PGWs).
		During stiff reheating phases with $\omega_{\rm re} > 1/3$, the high-frequency PGW spectrum is significantly enhanced, producing a distinct and potentially observable signature of the model.
		For suitable parameter values, the amplified PGW signal enters the sensitivity range of upcoming gravitational-wave detectors. \
		
		Overall, this work presents a unified and observationally consistent realization of holographic inflation in $f(R,T)$ gravity, constrained simultaneously by CMB measurements, swampland criteria, reheating physics, and PGW limits.
	
	\vspace{0.5cm}
\end{abstract}

%%%%%%%%%%%%%%%%%%%%%%%%%%%%%%%%%%%%%%%%%%%%%%%
\maketitle

%%%%%%%%%%%%%%%%%%%%%%%%%%%%%%%%%%%%%%%%%%%%%%%
%%%%%%%%%%%%%%%%%%%%%%%%%%%%%%%%%%%%%%%%%%%%%%%
%%%%%%%%%%%%%%%%%%%%%%%%%%%%%%%%%%%%%%%%%%%%%%%
%%%%%%%%%%%%%%%%%%%%%%%%%%%%%%%%%%%%%%%%%%%%%%%
% ============ Sec.I =======================================
%%%%%%%%%%%%%%%%%%%%%%%%%%%%%%%%%%%%%%%%%%%%%%%
%%%%%%%%%%%%%%%%%%%%%%%%%%%%%%%%%%%%%%%%%%%%%%%
%%%%%%%%%%%%%%%%%%%%%%%%%%%%%%%%%%%%%%%%%%%%%%%
%%%%%%%%%%%%%%%%%%%%%%%%%%%%%%%%%%%%%%%%%%%%%%%
\section{Introduction}\label{Sec_intro}

A central challenge in theoretical cosmology is to understand whether early-universe inflation can be generated by generalized energy densities arising in extensions of general relativity. Modified theories of gravity have therefore been widely investigated as potential frameworks capable of addressing open problems in both cosmology and astrophysics \cite{Myrzakulov:2025fQLm,Bhagat:2025WeylfQT,Bhardwaj:2025PDU,Samaddar:2025fQC,Naseer:2023fRT,Shamir:2022CompactfRT,Sharma:2025ANNfRT,Mohammadi:2025PLB,Kavya:2025fQ}. Among these proposals, the $f(R,T)$ model introduced in Ref.~\cite{Harko:2011kv} is particularly noteworthy, since the gravitational action depends not only on the Ricci scalar $R$ but also on the trace of the energy–momentum tensor $T$, leading to an explicit matter–geometry coupling. This feature has motivated a diverse range of applications, including compact stellar configurations~\cite{Pretel:2020oae,Pretel:2022qng}, dark-energy and dark-matter phenomenology~\cite{Bhatti:2016caw,Zaregonbadi:2016xna}, wormhole solutions~\cite{Moraes:2019pao}, gravitational-wave studies~\cite{Alves:2016iks}, and early-universe inflation~\cite{Bhattacharjee:2020jsf,Gamonal:2020itt,Sarkar:2022lir,Chen:2022dyq,Zhang:2021ppy}.

High-precision measurements of the cosmic microwave background (CMB), including the recent ACT~DR6 lensing data, reveal a mild preference for scalar spectral indices $n_s > 0.96$ at small scales. This upward shift challenges the predictions of the simplest single-field slow-roll models and has motivated the study of inflationary scenarios featuring additional degrees of freedom or modified background dynamics. Representative examples include models with noncanonical kinetic terms (such as k-inflation, tachyon, and DBI setups), multi-field constructions, Einstein--Gauss--Bonnet couplings, warm-inflation mechanisms, and holographic-inspired frameworks~\cite{Barenboim:2007ii,Franche:2010yj,Unnikrishnan:2012zu,Rezazadeh:2014fwa,Saaidi:2015kaa,Fairbairn:2002yp,Mukohyama:2002cn,Feinstein:2002aj,Padmanabhan:2002cp,berera1995warm,berera2000warm,hall2004scalar,Spalinski:2007dv,Bessada:2009pe,Weller:2011ey,Nazavari:2016yaa,Yogesh:2024zwi,Yogesh:2025hll,Yogesh:2025wak,Yogesh:2024vcl,Gangopadhyay:2022vgh,Khan:2022odn,maartens2000chaotic,golanbari2014brane,alexander2015dynamics,Tirandari:2017nzy,maeda2013stability,abolhasani2014primordial,Mohammadi:2020ake,Mohammadi:2020ctd,Mohammadi:2018oku,Mohammadi:2019dpu,Mohammadi:2018zkf,Mohammadi:2019qeu,Mohammadi:2020ftb,Mohammadi:2015upa,Mohammadi:2015jka}.

The nature of dark energy remains a central open question in cosmology~\cite{SupernovaSearchTeam:1998fmf,SupernovaCosmologyProject:1998vns,Boomerang:2000efg,Hanany:2000qf,Peebles:2002gy,PADMANABHAN2003235,Bamba:2012cp}. 
The holographic dark energy (HDE) paradigm~\cite{Cohen:1998zx,Li:2004rb,Hsu:2004ri,Horvat:2004vn} posits that the vacuum energy density scales as $\rho_{\rm HDE} \propto L^{-2}$, where $L$ is an infrared cutoff. Commonly adopted choices for $L$ include the Hubble radius, the particle or apparent horizon, the Ricci scale, and the Granda--Oliveros (GO) cutoff~\cite{Granda:2008dk,Granda:2008tm}. 
Although HDE is primarily employed to drive late-time cosmic acceleration~\cite{Li:2004rb,Nojiri:2019skr,Nojiri:2019itp,Nojiri:2005pu,Nojiri:2020wmh,Nojiri:2017opc,Nojiri:2021iko}, recent works have explored its potential role in the early Universe~\cite{Nojiri:2019kkp,Oliveros:2019rnq,Chakraborty_2020,Mohammadi:2021wde,Mohammadi:2021gvf,Mohammadi:2022vru}.

A particularly promising extension replaces the standard Bekenstein--Hawking entropy with the non-extensive Tsallis entropy $S \propto A^{\delta}$~\cite{Tsallis:1987eu,Tsallis_2013,Tsallis:1998ws,Lyra:1998wz,Wilk:1999dr,Lymperis:2018iuz}, yielding Tsallis holographic dark energy (THDE)~\cite{Rashki:2014noa,Tavayef:2018xwx}. 
The non-extensivity parameter $\delta$ introduces additional freedom, raising the intriguing possibility that an inflationary phase may emerge directly from a THDE density evaluated at a curvature-dependent cutoff (such as the GO scale), thereby eliminating the need for an \textit{ad hoc} inflaton potential. 
This approach offers an appealing unified framework in which early-universe acceleration is derived from the same holographic and thermodynamic principles that govern the late-time behaviour of dark energy.

Previous studies of inflation in $f(R,T)$ gravity~\cite{Bhattacharjee:2020jsf,Gamonal:2020itt,Sarkar:2022lir,Chen:2022dyq,Zhang:2021ppy} have typically relied on phenomenological inflaton potentials rather than deriving the inflationary dynamics from a thermodynamically motivated energy density. In particular, a fully self-consistent scenario in which a holographic energy density directly drives inflation, without invoking an explicit scalar-field potential, has remained largely unexplored. The present work addresses this gap by reconstructing the effective inflationary potential directly from a Tsallis holographic dark energy density evaluated at a curvature-dependent infrared cutoff, thereby providing a unified holographic description of both early- and late-time acceleration.

In this work, we develop an inflationary model in $f(R, T)$ gravity where the inflaton potential is replaced by the THDE energy density with the GO cutoff. Under slow-roll conditions, this approach yields a Hubble-dependent potential that can be mapped to $V(\phi)$ using standard reconstruction techniques. 
By comparing the model's predictions with the latest CMB measurements from ACT DR6, we constrain the relevant parameters. We also analyze the reheating phase, which is parametrized by the reheating temperature $T_{\rm re}$, the number of reheating e-folds $N_{\rm re}$, and the effective equation-of-state parameter $w_{\rm re}$\cite{Kofman:1994rk,Kofman:1997yn,Allahverdi:2010xz,Cook:2015vqa,Dai:2014jja,Munoz:2014eqa,Martin:2014nya,Rehagen:2015zma}. Furthermore, we study the spectrum of primordial gravitational waves (PGWs), whose high-frequency tail is enhanced during stiff reheating ($w_{\rm re} > 1/3$)\cite{Nakayama:2008wy,Boyle:2005se,Kuroyanagi:2009br} and is constrained by the CMB and BBN bound $\Delta N_{\rm eff} \leq 0.17$~\cite{Fields:2019pfx,Cyburt:2015mya}. This observational limit on the effective number of relativistic degrees of freedom imposes an additional lower bound on the reheating temperature, which is particularly stringent for stiff reheating. The resulting constraint on $T_{\rm re}$ translates into an upper limit on the number of inflationary e-folds.  
Finally, we examine the model's theoretical consistency against the Swampland conjectures~\cite{Obied:2018sgi,Garg:2018reu,Ooguri:2018wrx,PhysRevD.98.086004,Ooguri:2018Refined,Palti:2019Review,Mohammadi:2022fiv,Adhikari:2020xcg}, which impose restrictions on field excursions and potential gradients. Applied to the reconstructed THDE potential, these criteria typically require a matter--geometry coupling $\lambda \gtrsim \mathcal{O}(10^{2})$ to ensure sub-Planckian field evolution.

In summary, we present a unified inflationary framework where a Tsallis-type holographic energy density drives inflation within $f(R, T)$ gravity. 
To the best of our knowledge, this is the first comprehensive realization of THDE inflation embedded in $f(R,T)$ gravity that simultaneously incorporates CMB constraints, reheating dynamics, PGW phenomenology, and swampland bounds within a single consistent setup.
The remainder of the paper is organized as follows. Sec.~\ref{fRT_gravity} reviews $f(R,T)$ gravity. Sec.~\ref{inflation} summarizes the slow-roll framework and inflationary observables. In Sec.~\ref {pot_hde}, we introduce the Tsallis holographic potential with GO cutoff. Reheating and PGW constraints are examined in Sec.~\ref {reheating}. Sec.~\ref{reconstructing_potential} reconstructs the scalar field and potential and analyzes their swampland consistency. Finally, Sec.~\ref{conclusion} presents conclusions.

%%%%%%%%%%%%%%%%%%%%%%%%%%%%%%%%%%%%%%%%%%%%%%%
%%%%%%%%%%%%%%%%%%%%%%%%%%%%%%%%%%%%%%%%%%%%%%%
%%%%%%%%%%%%%%%%%%%%%%%%%%%%%%%%%%%%%%%%%%%%%%%
%%%%%%%%%%%%%%%%%%%%%%%%%%%%%%%%%%%%%%%%%%%%%%%
% ============ Sec.II =======================================
%%%%%%%%%%%%%%%%%%%%%%%%%%%%%%%%%%%%%%%%%%%%%%%
%%%%%%%%%%%%%%%%%%%%%%%%%%%%%%%%%%%%%%%%%%%%%%%
%%%%%%%%%%%%%%%%%%%%%%%%%%%%%%%%%%%%%%%%%%%%%%%
%%%%%%%%%%%%%%%%%%%%%%%%%%%%%%%%%%%%%%%%%%%%%%%
\section{$f(R,T)$ gravity theory}\label{fRT_gravity}

The $f(R,T)$ framework extends general relativity by allowing the gravitational Lagrangian to depend not only on the Ricci scalar $R$ but also on the trace of the energy--momentum tensor $T$. This additional dependence introduces an explicit nonminimal matter--geometry coupling, leading to deviations from Einsteinian dynamics and a richer phenomenology. The action is given by~\cite{Harko:2011kv}
\begin{equation}
	S = \frac{1}{2\kappa^2}\!\int d^4x\,\sqrt{-g}\, f(R,T) 
	+ \int d^4x\,\sqrt{-g}\,\mathcal{L}_m ,
\end{equation}
where $\mathcal{L}_m$ is the matter Lagrangian and $\kappa^2 = 8\pi G$.  
Varying this action with respect to the metric $g_{\mu\nu}$ yields
\begin{equation}\label{field_equation}
	\big( g_{\mu\nu}\Box - \nabla_\mu\nabla_\nu \big) f_{,R}
	+ f_{,R} R_{\mu\nu}
	- \frac{1}{2} g_{\mu\nu} f
	= \kappa^2 T_{\mu\nu} - f_{,T}\big( T_{\mu\nu} + \Theta_{\mu\nu} \big),
\end{equation}
where
\begin{equation}\label{Theta_definition}
	\Theta_{\mu\nu}
	\equiv g^{\alpha\beta} 
	\frac{\delta T_{\alpha\beta}}{\delta g^{\mu\nu}}
\end{equation}
encodes additional contributions arising from the matter--geometry coupling.
For a perfect fluid with energy density $\rho$, pressure $p$, and four-velocity $u_\mu$, the energy--momentum tensor takes the standard form
\begin{equation}
	T_{\mu\nu} = (\rho+p)\,u_\mu u_\nu - p\, g_{\mu\nu}.
\end{equation}
Choosing $\mathcal{L}_m = -p$ leads to~\cite{Harko:2011kv}
\begin{equation}\label{Theta_correct}
	\Theta_{\mu\nu} = -2T_{\mu\nu} - p\, g_{\mu\nu},
\end{equation}
and therefore $T = \rho - 3p$.
For cosmological applications we adopt the spatially flat FLRW metric
\begin{equation}
	ds^2 = dt^2 - a^2(t)\,\delta_{ij}\,dx^i dx^j ,
\end{equation}
with scale factor $a(t)$ and Hubble parameter $H=\dot{a}/a$.
To capture leading-order matter--geometry coupling effects while avoiding higher-derivative instabilities, we focus on the widely used linear model
\begin{equation}
	f(R,T) = R + \kappa^2\lambda T ,
\end{equation}
where $\lambda$ is a constant coupling parameter. For this choice, $f_{,R}=1$ and $f_{,T} = \kappa^2\lambda$, and Eq.~\eqref{field_equation} simplifies to
\begin{equation}
	G_{\mu\nu}
	= \kappa^2\!\left[(1+\lambda)\,T_{\mu\nu}
	+ \frac{\lambda}{2}(\rho-p)\, g_{\mu\nu}\right].
\end{equation}
Specializing to the FLRW background leads to the modified Friedmann equations
\begin{equation}\label{friedmann1}
	H^2 = \frac{\kappa^2}{3}
	\left[
	\left(1+\frac{3\lambda}{2}\right)\rho
	- \frac{\lambda}{2}p
	\right],
\end{equation}
\begin{equation}\label{friedmann2}
	-3H^2 - 2\dot{H}
	= \kappa^2\!\left[
	-\frac{\lambda}{2}\rho
	+\left(1+\frac{3\lambda}{2}\right)p
	\right].
\end{equation}
Combining these relations yields the modified Raychaudhuri equation
\begin{equation}\label{dHt}
	-2\dot{H} = \kappa^2 (1+\lambda)(\rho+p),
\end{equation}
which correctly reproduces the standard GR expression when $\lambda = 0$.
Because $f_{,T} \neq 0$, the matter sector is no longer conserved.  
Taking the time derivative of Eq.~\eqref{friedmann1} and using Eq.~\eqref{dHt} gives
\begin{equation}\label{conservation_correct}
	\dot{\rho} + 3H(\rho+p)
	= -\frac{3\lambda}{2}\dot{\rho}
	+ \frac{\lambda}{2}\dot{p}
	- 3\lambda H(\rho+p),
\end{equation}
which explicitly illustrates how the matter--geometry coupling modifies the usual continuity equation.  
These relations form the foundation for the slow-roll inflationary analysis developed in the next section.

%%%%%%%%%%%%%%%%%%%%%%%%%%%%%%%%%%%%%%%%%%%%%%%
%%%%%%%%%%%%%%%%%%%%%%%%%%%%%%%%%%%%%%%%%%%%%%%
%%%%%%%%%%%%%%%%%%%%%%%%%%%%%%%%%%%%%%%%%%%%%%%
%%%%%%%%%%%%%%%%%%%%%%%%%%%%%%%%%%%%%%%%%%%%%%%
% ============ Sec.III =======================================
%%%%%%%%%%%%%%%%%%%%%%%%%%%%%%%%%%%%%%%%%%%%%%%
%%%%%%%%%%%%%%%%%%%%%%%%%%%%%%%%%%%%%%%%%%%%%%%
%%%%%%%%%%%%%%%%%%%%%%%%%%%%%%%%%%%%%%%%%%%%%%%
%%%%%%%%%%%%%%%%%%%%%%%%%%%%%%%%%%%%%%%%%%%%%%%
\section{Slow-roll inflation}\label{inflation}

To study inflation in the $f(R,T)$ framework, we consider a canonical scalar field $\phi$ with Lagrangian
\begin{equation}\label{scalar_field_lagrangian}
	\mathcal{L}_\phi = \frac{1}{2}\partial_\mu\phi\,\partial^\mu\phi - V(\phi),
\end{equation}
which gives the energy density and pressure
\begin{align}\label{field_energy_density}
	\rho_\phi &= \tfrac{1}{2}\dot{\phi}^2 + V(\phi), \\
	p_\phi   &= \tfrac{1}{2}\dot{\phi}^2 - V(\phi).
\end{align}
Substituting these expressions into the modified Friedmann equations \eqref{friedmann1}--\eqref{dHt} leads to
\begin{align}
	H^2 &= \frac{\kappa^2}{3}\left[\tfrac{1}{2}(1+\lambda)\dot{\phi}^2 + (1+2\lambda)V(\phi)\right], \label{Fried1_field} \\
	-2\dot{H} &= \kappa^2(1+\lambda)\dot{\phi}^2. \label{Fried2_field}
\end{align}
Because of the matter--geometry coupling, the scalar-field dynamics is also modified.  
Using the generalized conservation equation \eqref{conservation_correct}, we obtain
\begin{equation}\label{field_eom}
	(1+\lambda)\ddot{\phi}
	+ 3H(1+\lambda)\dot{\phi}
	+ (1+2\lambda)V'(\phi)
	= 0,
\end{equation}
which shows that $\lambda$ changes both the friction term and the effective slope of the potential.
Inflation takes place in the slow-roll regime, where
\begin{equation*}
	|\dot{H}| \ll H^2, \qquad
	\dot{\phi}^2 \ll V(\phi), \qquad
	|\ddot{\phi}| \ll |H\dot{\phi}|.
\end{equation*}
Under these conditions, Eqs.~\eqref{Fried1_field}--\eqref{field_eom} simplify to
\begin{align}
	H^2 &= \frac{\kappa^2}{3}(1+2\lambda)V(\phi), \label{Fried1_field_sr} \\
	-2\dot{H} &= \kappa^2(1+\lambda)\dot{\phi}^2, \label{Fried2_field_sr} \\
	3H\dot{\phi} &= -\frac{1+2\lambda}{1+\lambda}\,V'(\phi). \label{field_eom_sr}
\end{align}
To describe departures from exact de Sitter expansion, we define the slow-roll hierarchy
\begin{equation}
	\epsilon_1 \equiv -\frac{\dot{H}}{H^2},
	\qquad
	\epsilon_{n+1} \equiv \frac{\dot{\epsilon}_n}{H\epsilon_n},
	\qquad (n \ge 1),
\end{equation}
and later use $\epsilon_1$ and $\epsilon_2$ to compute the inflationary observables.
In standard inflationary studies, the potential $V(\phi)$ is usually chosen as an external input to the model.  
In this approach, the potential is obtained directly from the Tsallis holographic energy density. Therefore, rather than being chosen arbitrarily, the potential emerges as a prediction of the model, allowing the inflationary dynamics to be expressed directly in terms of the Hubble parameter. This point will be addressed in detail in the next section.

%%%%%%%%%%%%%%%%%%%%%%%%%%%%%%%%%%%%%%%%%%%%%%%
%%%%%%%%%%%%%%%%%%%%%%%%%%%%%%%%%%%%%%%%%%%%%%%
%%%%%%%%%%%%%%%%%%%%%%%%%%%%%%%%%%%%%%%%%%%%%%%
%%%%%%%%%%%%%%%%%%%%%%%%%%%%%%%%%%%%%%%%%%%%%%%
% ============ Sec.IV =======================================
%%%%%%%%%%%%%%%%%%%%%%%%%%%%%%%%%%%%%%%%%%%%%%%
%%%%%%%%%%%%%%%%%%%%%%%%%%%%%%%%%%%%%%%%%%%%%%%
%%%%%%%%%%%%%%%%%%%%%%%%%%%%%%%%%%%%%%%%%%%%%%%
%%%%%%%%%%%%%%%%%%%%%%%%%%%%%%%%%%%%%%%%%%%%%%%
\section{THDE as the potential}\label{pot_hde}

During inflation, the total energy density is dominated by a single slowly evolving scalar field. In this regime, the kinetic term is subdominant, so that 
\begin{equation}
	\rho_{\phi} \simeq V(\phi).
\end{equation} 
In our approach, the inflaton potential is not assumed a priori; instead, it is derived directly from a generalized holographic energy density based on the non-extensive Tsallis entropy.
The holographic principle relates the number of degrees of freedom of a gravitating system to the area of its boundary~\cite{tHooft:1993dmi,Susskind:1994vu,Witten:1998qj,Bousso:2002ju}. Within Tsallis non-additive statistical mechanics~\cite{Tsallis:1987eu,Tsallis:1998ws}, the standard entropy--area relation is generalized as
\begin{equation}
	S = \gamma A^{\delta},
\end{equation}
where $\delta$ quantifies the degree of non-extensivity. Substituting this generalized entropy into the holographic construction yields the THDE:
\begin{equation}
	\rho_{\rm THDE} = B c^{2} L^{2\delta - 4},
\end{equation}
where $L$ denotes the infrared (IR) cutoff, $c^2>0$ is a constant, and $B \equiv \gamma (4 \pi)^\gamma$ collects the entropy prefactors. The Tsallis index $\delta$ determines the scaling of the energy density: $\delta = 1$ reproduces the standard holographic dark energy, while $\delta > 1$ introduces a steeper dependence on the IR scale, which in turn modifies the inflationary dynamics.
Thus, the Tsallis holographic energy density provides a physically motivated realization of inflation within $f(R,T)$ gravity and naturally incorporates UV/IR mixing through the Tsallis index $\delta$.
For the IR cutoff, we adopt the Granda--Oliveros (GO) prescription~\cite{Granda:2008dk,Granda:2008tm}:
\begin{equation}\label{GO}
	L^{-2} = \alpha H^2 + \beta \dot{H},
\end{equation}
where $\alpha$ and $\beta$ control the relative contributions of $H^2$ and $\dot{H}$. Negative values of $\beta$ effectively enhance the friction term in the background dynamics, an effect that is reflected in the slow-roll parameter $\epsilon_1$.
By identifying $V(\phi) = \rho_{\rm THDE}$ and using the slow-roll Friedmann relation from Sec.~\ref{inflation}, we obtain the key dynamical equation governing the inflationary evolution.
\begin{equation}\label{fried_hde}
	H^{2}= \frac{\kappa^{2}}{3}(1+2\lambda)\, Bc^{2}\big(\alpha H^{2}+\beta \dot{H}\big)^{\,2-\delta}.
\end{equation}
Equation~\eqref{fried_hde} can be rearranged to give the first slow-roll parameter
\begin{equation}\label{epsilon1}
	\epsilon_{1}= \frac{1}{\beta}\Bigl(\alpha - A H^{\xi}\Bigr),
\end{equation}
with  
\begin{equation}
	\xi\equiv\frac{2\delta-2}{2-\delta},
	\qquad
	A\equiv\left[\frac{3M_{p}^{2}}{Bc^{2}(1+2\lambda)}\right]^{1/(2-\delta)}.
\end{equation}
The second slow-roll parameter follows directly from its definition,
\begin{equation}\label{epsilon2}
	\epsilon_{2}=\frac{\dot{\epsilon}_{1}}{H\epsilon_{1}}
	= \frac{\xi}{\beta}\, A H^{\xi}.
\end{equation}
The number of e-folds accumulated between horizon exit and the end of inflation is
\begin{equation}\label{efold_correct}
	N= \int_{t_{\star}}^{t_{e}}H\,dt
	= \int_{H_{\star}}^{H_{e}}\frac{H}{\dot{H}}\,dH ,
\end{equation}
where we define $N$ as the absolute number of e-folds between horizon exit and the end of inflation, so that $N>0$ even though $H$ decreases during inflation.  
The end of inflation is fixed by $\epsilon_{1}(H_{e})=1$, which implies $A H_{e}^{\xi}=\alpha-\beta$.  
Using Eq.~\eqref{epsilon1}, Eq.~\eqref{efold_correct} integrates to
\begin{equation}\label{AH_s}
	A H_{\star}^{\xi} =
	\frac{\alpha(\alpha-\beta)\,\exp\!\bigl(\tfrac{\alpha\xi}{\beta}N\bigr)}
	{\beta+(\alpha-\beta)\,\exp\!\bigl(\tfrac{\alpha\xi}{\beta}N\bigr)}.
\end{equation}
The inflationary observables follow from the standard slow-roll relations:
\begin{align}
	n_{s} & =1-(2\epsilon_{1}+\epsilon_{2})
	= 1-\frac{2\alpha}{\beta}
	- \frac{\xi-2}{\beta}\,A H^{\xi},\label{ns_slowroll}\\[1mm]
	r     & = \frac{16}{1+\lambda}\,\epsilon_{1}
	= \frac{16}{\beta(1+\lambda)}\bigl(\alpha-AH^{\xi}\bigr),\label{r_slowroll}
\end{align}
evaluated at horizon crossing using Eq.~\eqref{AH_s}.  
These expressions show how the four parameters $(\alpha,\beta,\delta,\lambda)$ determine the inflationary predictions. 
In particular, the Tsallis index $\delta$ enters through the combination $\xi(\delta)$, which controls the $H$-dependence of the slow-roll parameters and therefore the effective steepness of the reconstructed potential.  
The matter--geometry coupling $\lambda$ modifies the effective kinetic normalization via the factors $(1+\lambda)$ and $(1+2\lambda)$ in the background equations, and it suppresses the tensor amplitude through the explicit $(1+\lambda)^{-1}$ factor in Eq.~\eqref{r_slowroll}.
In what follows, we fix $N=65$, consistent with a standard thermal history, and compare the predictions of Eqs.~\eqref{ns_slowroll} and~\eqref{r_slowroll} with the combined P-ACT-LB constraints based on \textit{Planck}~2018 and ACT~DR6 temperature and polarization measurements.

\begin{figure}[htbp]
	\centering 
	\subfigure[\label{rns_beta}]{\includegraphics[width = 0.4\linewidth]{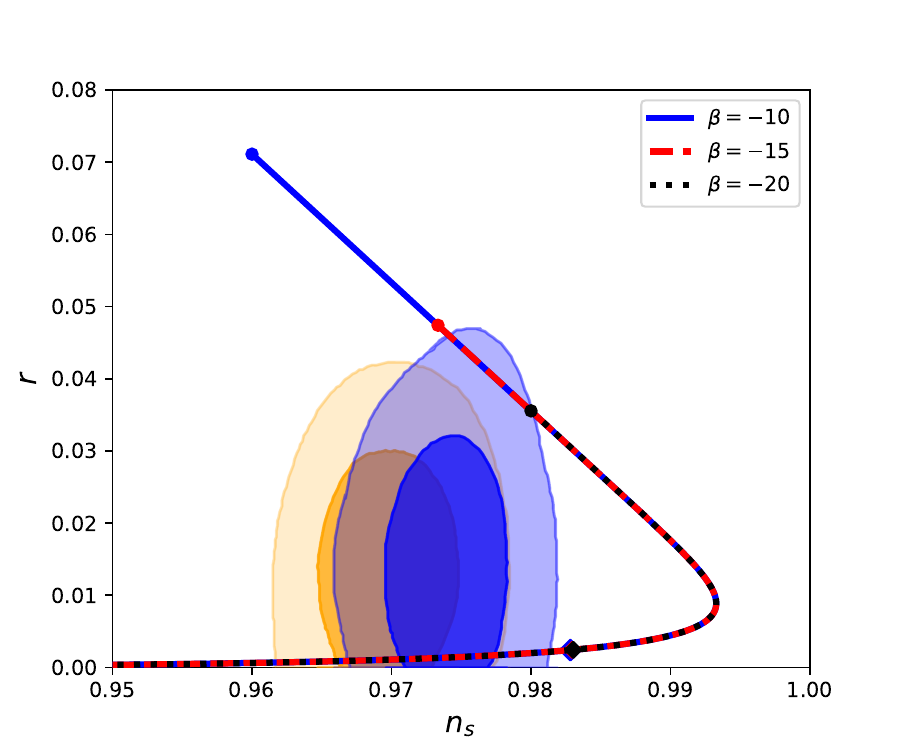}}
	\subfigure[\label{rns_delta}]{\includegraphics[width = 0.4\linewidth]{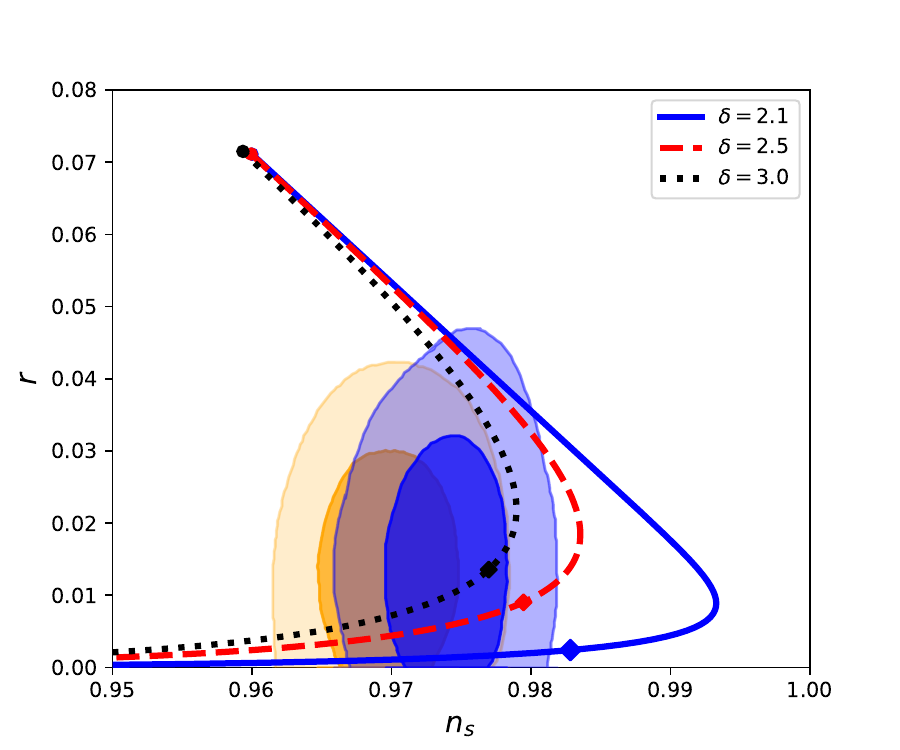}}
	\subfigure[\label{rns_lambda}]{\includegraphics[width = 0.4\linewidth]{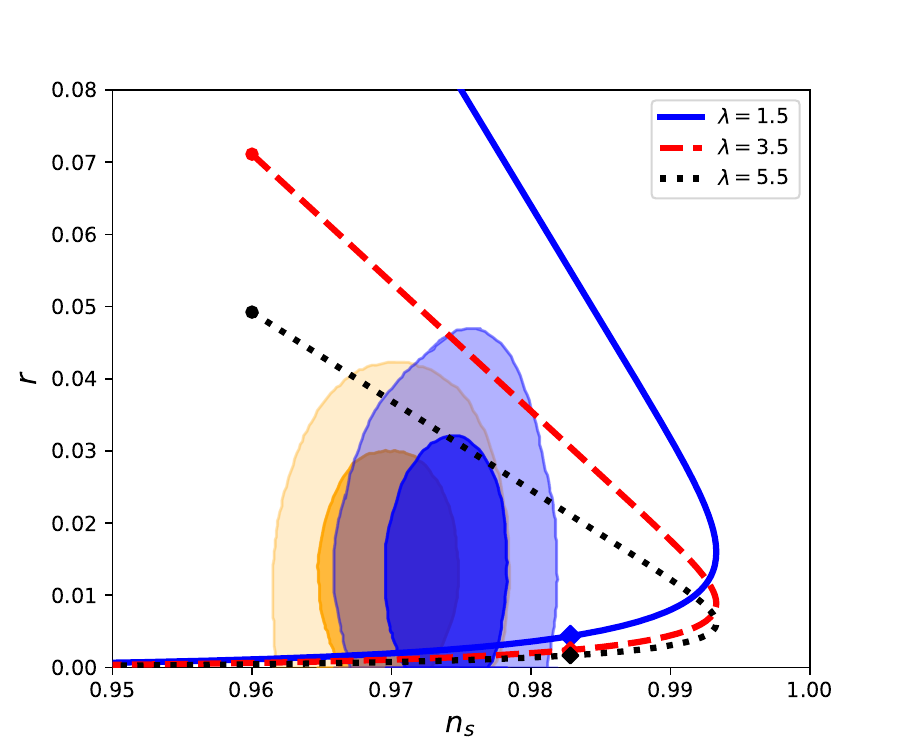}}
	\caption{\label{rns}
		Trajectories in the $(n_s,r)$ plane obtained by varying:
		(a)~$\beta$ at fixed $(\delta,\lambda)=(2.1,3.5)$,
		(b)~$\delta$ at fixed $(\beta,\lambda)=(-10,3.5)$,
		and (c)~$\lambda$ at fixed $(\beta,\delta)=(-10,2.1)$.
		Along each curve $\alpha$ is scanned over $-0.2\le\alpha\le 0.2$ with $N=65$.
		The orange band shows the \textit{Planck}~2018 constraints,
		and the blue regions correspond to the $68\%$ and $95\%$
		confidence intervals of the combined P-ACT-LB data set.
	}
\end{figure}

Figures~\ref{rns} summarize the resulting trajectories in the $(n_{s},r)$ plane.  
In all panels, the parameter $\alpha$ is scanned over the interval $-0.2\le\alpha\le0.2$, while the background evolution is determined by Eqs.~\eqref{fried_hde}--\eqref{epsilon2}.\\
Figure~\ref{rns}(a) shows the effect of varying~$\beta$ at fixed $(\delta,\lambda)=(2.1,3.5)$.  
Since $\epsilon_{1}$ depends linearly on $1/\beta$, more negative $\beta$ values reduce~$\epsilon_{1}$ and shift the curves toward smaller~$r$.\\
Figure~\ref{rns}(b) keeps $(\beta,\lambda)=(-10,3.5)$ fixed and varies $\delta$.  
Because $\xi(\delta)$ enters $\epsilon_{2}$, larger $\delta$ slightly lowers~$n_{s}$ and changes the curvature of the trajectories.\\
Figure~\ref{rns}(c) shows the effect of the matter--geometry coupling at $(\beta,\delta)=(-10,2.1)$.  
Increasing $\lambda$ suppresses $r$ through Eq.~\eqref{r_slowroll}, moving the predictions deeper into the region favored by the P-ACT-LB analysis.
%%%%%%%%%%%%%%%%%%%%%%%%%%%%%%%%%%%%
%%%%%%%%%%%%%%%%%%%%%%%%%%%%%%%%%%%%

\begin{figure}[htbp]
	\centering 
	\subfigure[\label{param_space_ab}]{\includegraphics[width = 0.4 \linewidth]{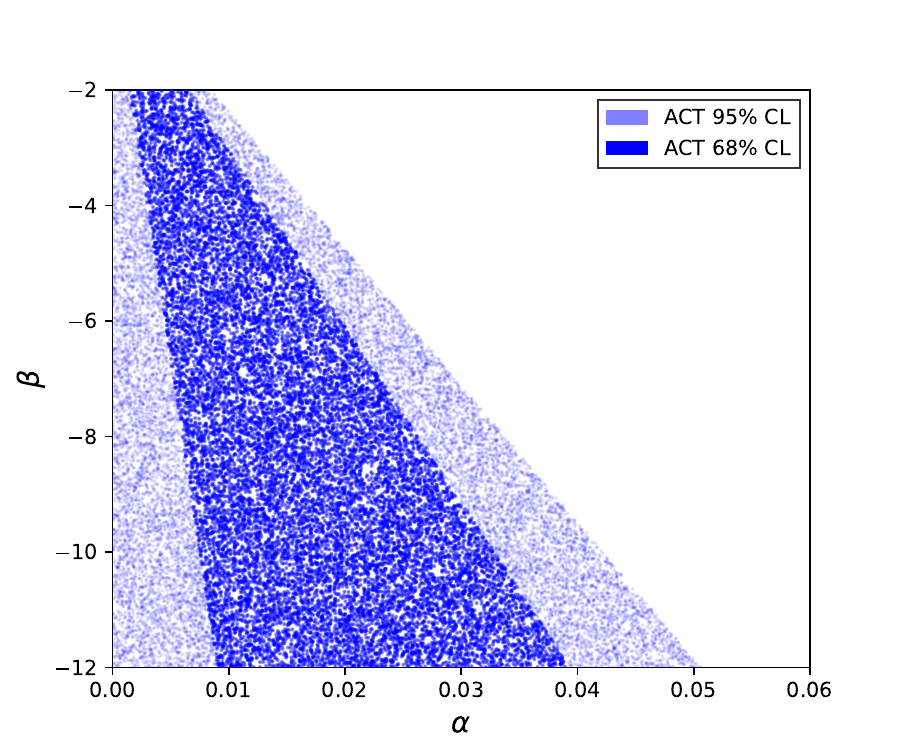}}
	\subfigure[\label{param_space_ad}]{\includegraphics[width = 0.4 \linewidth]{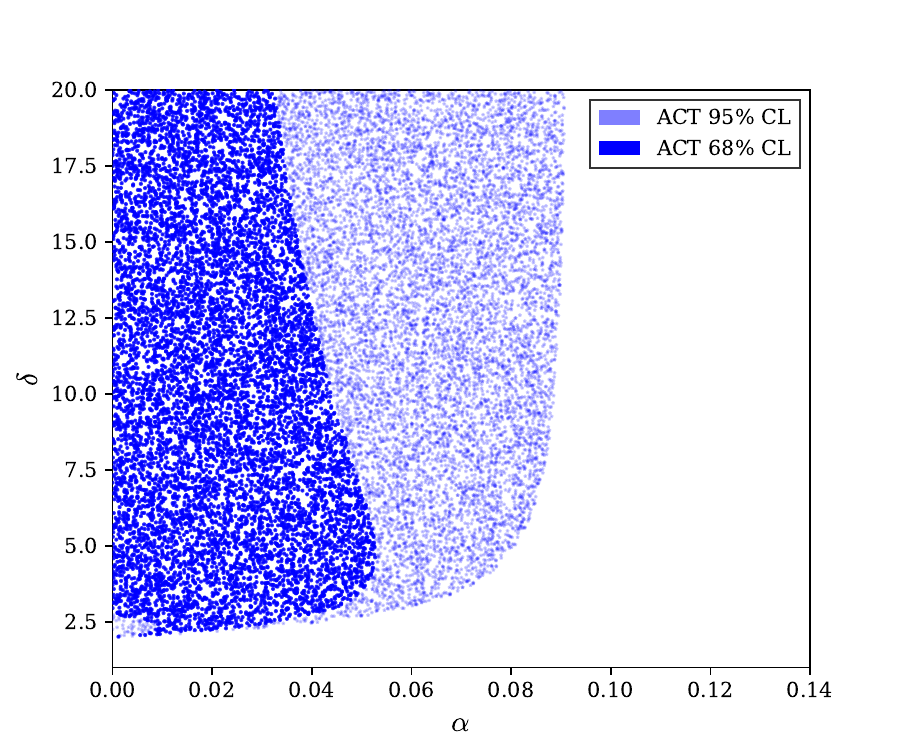}}
	\subfigure[\label{param_space_bd}]{\includegraphics[width = 0.4 \linewidth]{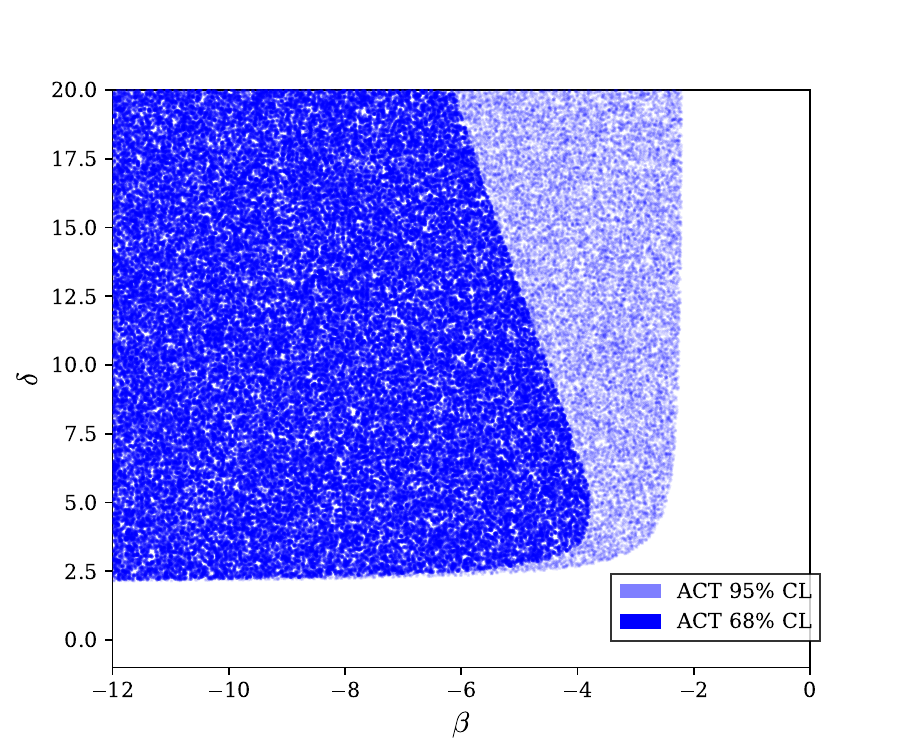}}
	\caption{\label{param_space}
		Parameter-space regions consistent with the combined P-ACT-LB constraints.
		(a) $(\alpha,\beta)$ plane for $\delta=2.5$;
		(b) $(\alpha,\delta)$ plane for $\beta=-10$;
		(c) $(\beta,\delta)$ plane for $\alpha=0.02$.
		In all cases $\lambda=5$ and $N=65$ are fixed.
		These slices highlight the combined roles of the GO parameters and the Tsallis index in reproducing the observed $(n_s,r)$ values.
	}
\end{figure}

To further explore the parameter space, Fig.~\ref{param_space} shows slices of the $(\alpha,\beta,\delta)$ space for which Eqs.~\eqref{ns_slowroll} and~\eqref{r_slowroll} give values of $(n_s,r)$ inside the $68\%$ P-ACT-LB region.  
In all three panels we fix $\lambda=5$ and $N=65$.\\
Figure~\ref{param_space}(a) displays the $(\alpha,\beta)$ plane for $\delta=2.5$.  
Viable points lie in the region of negative~$\beta$, consistent with the requirement $\epsilon_{1}>0$ from Eq.~\eqref{epsilon1}.\\
Figure~\ref{param_space}(b) shows the $(\alpha,\delta)$ plane for $\beta=-10$.  
Observational consistency restricts~$\delta$ to a relatively narrow band.\\
Figure~\ref{param_space}(c) presents the $(\beta,\delta)$ plane for $\alpha=0.02$.  
This panel shows the correlated ranges of~$\beta$ and~$\delta$ required to reproduce the observed tilt.
%%%%%%%%%%%%%%%%%%%%%%%%%%%%%%%%%%%%
%%%%%%%%%%%%%%%%%%%%%%%%%%%%%%%%%%%%

\begin{figure}[htbp]
	\centering
	\includegraphics[width = 0.4 \linewidth]{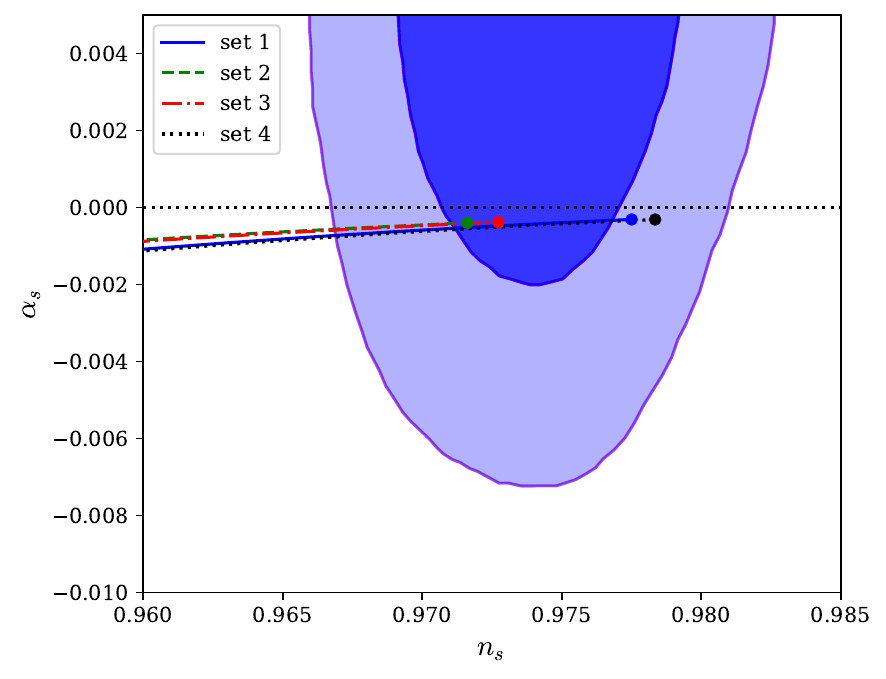}
	\caption{\label{running}
		Running of the scalar spectral index $\alpha_s$ versus $n_s$
		for four representative parameter sets $(\alpha,\beta,\delta)$:
		set1 = $(0.01,-11,2.5)$,
		set2 = $(0.03,-10,5)$,
		set3 = $(0.02,-8,4.0)$,
		and set4 = $(0.005,-9,2.5)$,
		with $\lambda=5$.
		Along each trajectory, $N$ varies, and the marked points correspond to $N=65$.
		In the observationally relevant region, The resulting running is small and negative of the order of $\mathcal{O}(10^{-3})$, consistent with the combined P-ACT-LB bounds.
	}
\end{figure}

The scale dependence of the scalar spectrum is further illustrated in Fig.~\ref{running}, which shows the running $\alpha_s = d n_s/d\ln k$ as a function of $n_s$ for three representative sets $(\alpha,\beta,\delta)$ with $\lambda=5$.  
Along each trajectory, the number of e-folds varies, and the marked points correspond to $N=65$.  
Near horizon crossing, the running is small and negative, $\alpha_s \sim \mathcal{O}(10^{-3})$.  
This is consistent with the P-ACT-LB bounds and a stable slow-roll evolution.
%%%%%%%%%%%%%%%%%%%%%%%%%%%%%%%%%%%%
%%%%%%%%%%%%%%%%%%%%%%%%%%%%%%%%%%%%

Finally, Table~\ref{table_rns} lists a set of representative parameter choices and the corresponding observables.  
For all entries, we fix $N=65$ and compute $(n_s,r)$ from Eqs.~\eqref{ns_slowroll} and~\eqref{r_slowroll}.  
The scalar tilt lies in the interval $0.9687\lesssim n_s\lesssim 0.973$, while the tensor-to-scalar ratio spans $0.0059\lesssim r\lesssim 0.024$, well inside the $68\%$ P-ACT-LB region.  
The combination $Bc^2$ decreases rapidly as $\delta$ increases.  
This demonstrates the strong sensitivity of the inflationary energy scale (ES) to the non-extensive index $\delta$.  
Comparison of the blocks with $\lambda=3$ and $\lambda=8$ also shows the strong suppressive effect of the $f(R,T)$ coupling on~$r$, as expected from Eq.~\eqref{r_slowroll}.  
Taken together, these results demonstrate that a Tsallis holographic potential in $f(R,T)$ gravity can reproduce the current CMB constraints on $(n_s,r)$ over a broad region of parameter space.
%%%%%%%%%%%%%%%%%%%%%%%%%%%%%%%%%%%%
\begin{table}[t]
	\centering
	\setlength{\tabcolsep}{10pt}
	\renewcommand{\arraystretch}{1.2}
	\caption{Representative parameter sets for the Tsallis holographic inflation model.  
		Each row lists $(\lambda,\alpha,\beta,\delta)$ together with the resulting scalar tilt $n_s$, 
		the tensor-to-scalar ratio $r$, the combination $Bc^2$, and the inflationary energy scale (ES), all evaluated at $N=65$.  
		All entries lie within the $68\%$ P-ACT-LB region.}
	\label{table_rns}
	\begin{tabular}{cccccccc}
		\hline\hline
		$\lambda$   & $\alpha$   & $\beta$   & $\delta$   & $n_s$    & $r$         & $Bc^2$      & ES [$10^{-3}$]          \\
		\hline
		3.0 & 0.01 & -12.0 & 3.5 & 0.9728 & $1.828 \times 10^{-2}$ & $6.442 \times 10^{-26}$ & $4.017$ \\
		3.0 & 0.01 & -12.0 & 4.5 & 0.9712 & $2.204 \times 10^{-2}$ & $9.090 \times 10^{-36}$ & $4.276$ \\
		3.0 & 0.01 & -10.0 & 3.5 & 0.9727 & $1.797 \times 10^{-2}$ & $4.810 \times 10^{-26}$ & $4.000$ \\
		3.0 & 0.01 & -10.0 & 4.5 & 0.9711 & $2.173 \times 10^{-2}$ & $5.669 \times 10^{-36}$ & $4.261$ \\
		3.0 & 0.01 & -8.0  & 3.5 & 0.9725 & $1.751 \times 10^{-2}$ & $3.343 \times 10^{-26}$ & $3.974$ \\
		3.0 & 0.01 & -8.0  & 4.5 & 0.9710 & $2.126 \times 10^{-2}$ & $3.165 \times 10^{-36}$ & $4.238$ \\
		3.0 & 0.02 & -12.0 & 3.5 & 0.9721 & $1.675 \times 10^{-2}$ & $5.842 \times 10^{-26}$ & $3.931$ \\
		3.0 & 0.02 & -12.0 & 4.5 & 0.9707 & $2.049 \times 10^{-2}$ & $8.347 \times 10^{-36}$ & $4.199$ \\
		3.0 & 0.02 & -10.0 & 3.5 & 0.9718 & $1.617 \times 10^{-2}$ & $4.263 \times 10^{-26}$ & $3.896$ \\
		3.0 & 0.02 & -10.0 & 4.5 & 0.9705 & $1.989 \times 10^{-2}$ & $5.100 \times 10^{-36}$ & $4.168$ \\
		3.0 & 0.02 & -8.0  & 3.5 & 0.9712 & $1.532 \times 10^{-2}$ & $2.857 \times 10^{-26}$ & $3.844$ \\
		3.0 & 0.02 & -8.0  & 4.5 & 0.9702 & $1.902 \times 10^{-2}$ & $2.754 \times 10^{-36}$ & $4.121$ \\
		3.0 & 0.03 & -12.0 & 3.5 & 0.9712 & $1.532 \times 10^{-2}$ & $5.249 \times 10^{-26}$ & $3.844$ \\
		3.0 & 0.03 & -12.0 & 4.5 & 0.9702 & $1.902 \times 10^{-2}$ & $7.589 \times 10^{-36}$ & $4.121$ \\
		3.0 & 0.03 & -10.0 & 3.5 & 0.9707 & $1.450 \times 10^{-2}$ & $3.728 \times 10^{-26}$ & $3.791$ \\
		3.0 & 0.03 & -10.0 & 4.5 & 0.9698 & $1.817 \times 10^{-2}$ & $4.523 \times 10^{-36}$ & $4.075$ \\
		3.0 & 0.03 & -8.0  & 3.5 & 0.9697 & $1.333 \times 10^{-2}$ & $2.392 \times 10^{-26}$ & $3.713$ \\
		3.0 & 0.03 & -8.0  & 4.5 & 0.9692 & $1.696 \times 10^{-2}$ & $2.344 \times 10^{-36}$ & $4.005$ \\
		8.0 & 0.01 & -12.0 & 3.5 & 0.9728 & $8.125 \times 10^{-3}$ & $2.653 \times 10^{-26}$ & $3.218$ \\
		8.0 & 0.01 & -12.0 & 4.5 & 0.9712 & $9.796 \times 10^{-3}$ & $3.743 \times 10^{-36}$ & $3.425$ \\
		8.0 & 0.01 & -10.0 & 3.5 & 0.9727 & $7.986 \times 10^{-3}$ & $1.980 \times 10^{-26}$ & $3.204$ \\
		8.0 & 0.01 & -10.0 & 4.5 & 0.9711 & $9.656 \times 10^{-3}$ & $2.334 \times 10^{-36}$ & $3.413$ \\
		8.0 & 0.01 & -8.0  & 3.5 & 0.9725 & $7.780 \times 10^{-3}$ & $1.377 \times 10^{-26}$ & $3.183$ \\
		8.0 & 0.01 & -8.0  & 4.5 & 0.9710 & $9.447 \times 10^{-3}$ & $1.303 \times 10^{-36}$ & $3.395 $ \\
		8.0 & 0.02 & -12.0 & 3.5 & 0.9721 & $7.446 \times 10^{-3}$ & $2.406 \times 10^{-26}$ & $3.149$ \\
		8.0 & 0.02 & -12.0 & 4.5 & 0.9707 & $9.107 \times 10^{-3}$ & $3.437 \times 10^{-36}$ & $3.364$ \\
		8.0 & 0.02 & -10.0 & 3.5 & 0.9718 & $7.186 \times 10^{-3}$ & $1.755 \times 10^{-26}$ & $3.121$ \\
		8.0 & 0.02 & -10.0 & 4.5 & 0.9705 & $8.842 \times 10^{-3}$ & $2.100 \times 10^{-36}$ & $3.339$ \\
		8.0 & 0.02 & -8.0  & 3.5 & 0.9712 & $6.808 \times 10^{-3}$ & $1.177 \times 10^{-26}$ & $3.079$ \\
		8.0 & 0.02 & -8.0  & 4.5 & 0.9702 & $8.453 \times 10^{-3}$ & $1.134 \times 10^{-36}$ & $3.302$ \\
		8.0 & 0.03 & -12.0 & 3.5 & 0.9712 & $6.808 \times 10^{-3}$ & $2.161 \times 10^{-26}$ & $3.079$ \\
		8.0 & 0.03 & -12.0 & 4.5 & 0.9702 & $8.453 \times 10^{-3}$ & $3.125 \times 10^{-36}$ & $3.302$ \\
		8.0 & 0.03 & -10.0 & 3.5 & 0.9707 & $6.444 \times 10^{-3}$ & $1.535 \times 10^{-26}$ & $3.037$ \\
		8.0 & 0.03 & -10.0 & 4.5 & 0.9698 & $8.077 \times 10^{-3}$ & $1.862 \times 10^{-36}$ & $3.264$ \\
		8.0 & 0.03 & -8.0  & 3.5 & 0.9697 & $5.926 \times 10^{-3}$ & $9.848 \times 10^{-27}$ & $2.974$ \\
		8.0 & 0.03 & -8.0  & 4.5 & 0.9692 & $7.536 \times 10^{-3}$ & $9.653 \times 10^{-37}$ & $3.208$ \\
		\hline
	\end{tabular}
\end{table}
%%%%%%%%%%%%%%%%%%%%%%%%%%%%%%%%%%%%

%%%%%%%%%%%%%%%%%%%%%%%%%%%%%%%%%%%%%%%%%%%%%%%
%%%%%%%%%%%%%%%%%%%%%%%%%%%%%%%%%%%%%%%%%%%%%%%
%%%%%%%%%%%%%%%%%%%%%%%%%%%%%%%%%%%%%%%%%%%%%%%
%%%%%%%%%%%%%%%%%%%%%%%%%%%%%%%%%%%%%%%%%%%%%%%
% ============ Sec.V =======================================
%%%%%%%%%%%%%%%%%%%%%%%%%%%%%%%%%%%%%%%%%%%%%%%
%%%%%%%%%%%%%%%%%%%%%%%%%%%%%%%%%%%%%%%%%%%%%%%
%%%%%%%%%%%%%%%%%%%%%%%%%%%%%%%%%%%%%%%%%%%%%%%
%%%%%%%%%%%%%%%%%%%%%%%%%%%%%%%%%%%%%%%%%%%%%%%
\clearpage
\section{Reheating and Primordial Gravitational Waves}\label{reheating}

The accelerated expansion during inflation drives the Universe into an extremely cold state in which ordinary particles are almost absent. 
In order to recover the thermal conditions required for the onset of the hot Big Bang phase, a reheating era must take place to heat up the universe and provide a smooth transition to the radiation-dominant phase. 
During this stage, the energy stored in the inflaton field is transferred into relativistic species, eventually leading to a thermal bath in equilibrium. This transition is described by a few macroscopic quantities that capture the post-inflationary evolution. Universe~\cite{Albrecht:1982mp,Traschen:1990sw,Shtanov:1994ce,Kofman:1997yn,Bassett:2005xm,Allahverdi:2010xz,Martin:2014nya,Rehagen:2015zma,Cook:2015vqa,Yogesh:2024iip,Yogesh:2024mpa,Adhikari:2019uaw}.\\

The reheating phase is often studied indirectly through the reheating temperature, where inflationary observable parameters play a key role. Direct observational constraints on reheating are very limited, which makes its study challenging. The reheating phase usually described in terms of three parameters:
%The background evolution between the end of inflation and the onset of radiation domination can be described in terms of:
(i) the number of reheating $e$-folds $N_{\mathrm{re}}$,  
(ii) the reheating temperature $T_{\mathrm{re}}$,  
and (iii) the effective equation-of-state parameter $\omega_{\mathrm{re}}$ during this epoch.  
Assuming $\omega_{\mathrm{re}}$ remains constant during the reheating, the energy density evolves as
\begin{equation}
	\rho_{\mathrm{re}} = \rho_{e}\,
	\exp\!\left[-3(1+\omega_{\mathrm{re}}) N_{\mathrm{re}}\right],
\end{equation}
where $\rho_{e} = 3 M_{p}^{2} H_{e}^{2}$ is the total energy density at the end of inflation. 
After thermalization, the energy density is given by the standard expression $\rho_{\mathrm{re}} = \frac{\pi^{2}}{30}\, g_{\star\mathrm{re}}\, T_{\mathrm{re}}^{4},$ with $g_{\star\mathrm{re}}$ denoting the effective number of relativistic degrees of freedom. Equating these two expressions, one can find the the reheating e-folds $N_{\mathrm{re}}$ in terms of the reheating temperature $T_{\mathrm{re}}$ as
\begin{equation}\label{eq:Nre_standard}
	N_{\mathrm{re}} = -\frac{1}{3(1+\omega_{\mathrm{re}})}
	\ln\!\left(\frac{\pi^{2} g_{\star\mathrm{re}}}{90\, M_{p}^{2} H_{e}^{2}}\right)
	- \frac{4}{3(1+\omega_{\mathrm{re}})} \ln T_{\mathrm{re}},
\end{equation}
independent of the inflationary model.
%which holds independently of the inflationary model. 
The entropy is proportional to the temperature, i.e. $s \propto g_{\star s} T^{3}$, and the comoving entropy is conserved, $a^3 g_{\star s} T^{3} = \rm{cte}$. Applying this conservation between the reheating epoch and the present day yields the reheating temperature in terms of both reheating e-folds and inflationary e-folds
\begin{equation}\label{eq:Tre_standard}
	T_{\mathrm{re}} =
	\left( \frac{43}{11\, g_{\star s,\mathrm{re}}} \right)^{1/3}
	\frac{H_{k}\, T_{0}}{k_{\star}}\,
	\exp\!\left[-(N_{k} + N_{\mathrm{re}})\right],
\end{equation}
where $T_{0}$ is today’s CMB temperature, $H_{k}$ is the Hubble scale at the horizon exit of the pivot mode $k_{\star}$, and $N_{k}$ is the number of $e$-folds from $k_{\star}$ exit to the end of inflation.  
%\textcolor{blue}{Throughout the numerical analysis we fix $g_{\star{\rm re}} = g_{\star s,{\rm re}} = 106.75$, in agreement with the relativistic degrees of freedom of the Standard Model at high temperature.}
Combining Eqs.~\eqref{eq:Nre_standard} and~\eqref{eq:Tre_standard} provides a closed expression for the reheating temperature,
\begin{equation}\label{eq:Tre_closed}
	T_{\mathrm{re}} =
	\left[
	\left(\frac{11\, g_{\star s,\mathrm{re}}}{43}\right)^{1/3}
	\left(\frac{90\, M_{p}^{2} H_{e}^{2}}{\pi^{2} g_{\star\mathrm{re}}}\right)^{\!\frac{1}{3(1+\omega_{\mathrm{re}})}}
	\frac{k_{\star}}{H_{k} T_{0}}\, e^{N_{k}}
	\right]^{\!\frac{3(1+\omega_{\mathrm{re}})}{1 - 3\omega_{\mathrm{re}}}}.
\end{equation}
The reheating temperature can vary over many order of magnitude as there is a wide acceptable range for it. The general lower bound is BBN temperature $T_{BBN} \simeq 4\,\mathrm{MeV}$~\cite{Kawasaki:2000en,Kawasaki:2005nd,Dai:2014jja}, to ensure the radiation-dominant phase. The general upper bound is set by GUT as $10^{16}\,{\rm GeV}$ to avoid excessive reheating that could disrupt inflationary predictions. Additionally, the PGWs can provide a tighter limit on the reheating temperature in certain scenarios \cite{Boyle:2005se,Watanabe:2006qe,Saikawa:2018rcs,Caprini:2018mtu,Figueroa:2019paj,Bernal:2019lpc,Bernal:2020ywq,Bhattacharya:2020lhc,Bhattacharya:2019bvk}. The PGW, originated from the tensor perturbation during inflation, re-enter the horizon, propagate and evolve through the post-inflationary era, which can enhance the effective relativistic degrees of freedom $\Delta N_{\rm eff}$. Studies determined that this effect is become pronounced during stiff reheating ($w_{\rm re} > 1/3$) \cite{Boyle:2005se,Watanabe:2006qe,Saikawa:2018rcs,Caprini:2018mtu}, influencing the reheating dynamics and requiring careful analysis to align with observational bounds on $\Delta N_{\rm eff} \le 0.17$; as inferred from the combined P-ACT-LB analysis.
%Tensor perturbations generated during inflation form a stochastic background of PGWs. 
%Modes which re-enter the horizon before the onset of radiation domination can gain additional enhancement if the post-inflationary expansion is stiffer than radiation~\cite{Boyle:2005se,Watanabe:2006qe,Saikawa:2018rcs,Caprini:2018mtu}. 
%The cumulative contribution of these high-frequency modes to the radiation energy density is constrained by the bounds on the excess-radiation parameter $\Delta N_{\mathrm{eff}} \le 0.17,$
%\textcolor{blue}{as inferred from the combined P-ACT-LB analysis.}
The present GW energy density must satisfy the integral condition~\cite{Haque:2021dha,Chakraborty:2023ocr,Mohammadi:2025gbu}
\begin{equation}\label{eq:PGW_integral}
	\int_{k_{\mathrm{re}}}^{k_{e}}
	\frac{dk}{k}\, \Omega_{\mathrm{GW}}^{(0)}(k) h^{2}
	\;\le\;
	\frac{7}{8}
	\left(\frac{4}{11}\right)^{4/3}
	\Omega_{\gamma}^{(0)} h^{2}\,
	\Delta N_{\mathrm{eff}},
\end{equation}
where $k_{e} = a_{e} H_{e}$ corresponds to modes exiting the horizon at the end of inflation and $k_{\mathrm{re}} = a_{\mathrm{re}} H_{\mathrm{re}}$ is associated with modes re-entering at the end of reheating  \cite{Nakayama:2008wy,Bernal:2020ywq}.
\begin{equation}\label{eq:PGW_spectrum}
	\Omega_{\mathrm{GW}}^{(0)}(k) h^{2} \propto
	\left(\frac{k}{k_{\mathrm{re}}}\right)^{\frac{6\omega_{\mathrm{re}} - 2}{1 + 3\omega_{\mathrm{re}}}},
\end{equation}
which steepens the high-frequency slope of the PGW spectrum. 
Imposing Eq.~\eqref{eq:PGW_integral} leads to a lower limit on the reheating temperature,
\begin{equation}\label{eq:TreGW_standard}
	T_{\mathrm{re}} \ge T_{\mathrm{re}}^{\mathrm{GW}} =
	\left( \frac{90\, H_{e}^{2} M_{p}^{2}}{\pi^{2} g_{\star\mathrm{re}}} \right)^{1/4}
	\bigg[
	\frac{\Omega_{R}^{(0)} h^{2}}{5.61\times 10^{-6}\, \Delta N_{\mathrm{eff}}}
	\frac{H_{e}^{2}}{12\pi M_{p}^{2}}
	\frac{(1+3\omega_{\mathrm{re}})^{2}}{3\omega_{\mathrm{re}} - 1}
	\mu(\omega_{\mathrm{re}})
	\bigg]^{\frac{3(1+\omega_{\mathrm{re}})}{4(3\omega_{\mathrm{re}} - 1)}},
\end{equation}
where the quantity $\mu(\omega_{\mathrm{re}})$ is defined as 
\begin{equation}
	\mu(\omega_{\mathrm{re}}) =
	(1+3\omega_{\mathrm{re}})^{\frac{4}{1+3\omega_{\mathrm{re}}}}
	\Gamma^{2}\!\left(\frac{5+3\omega_{\mathrm{re}}}{2+6\omega_{\mathrm{re}}}\right)
\end{equation}
which is of the order of $\mathcal{O}(1)$ parameter. This constraint on the reheating parameter becomes efficient for $\omega_{\rm re} \gtrsim 0.6$, and provides a new lower limit on the reheating $T_{\rm re}$, and results in a constraint on the energy scale of inflation, and consequently constrains the parameters of the model.

\begin{figure}[htbp]
	\centering 
	\subfigure[\label{Tre01}]{\includegraphics[width=0.4\linewidth]{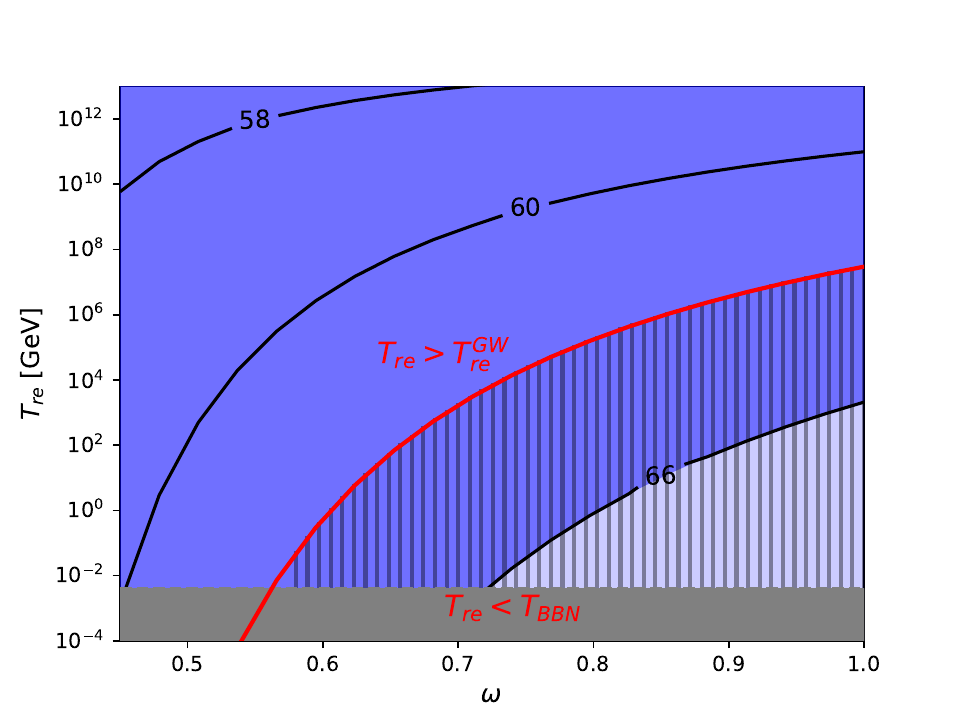}}
	\subfigure[\label{Tre02}]{\includegraphics[width=0.4\linewidth]{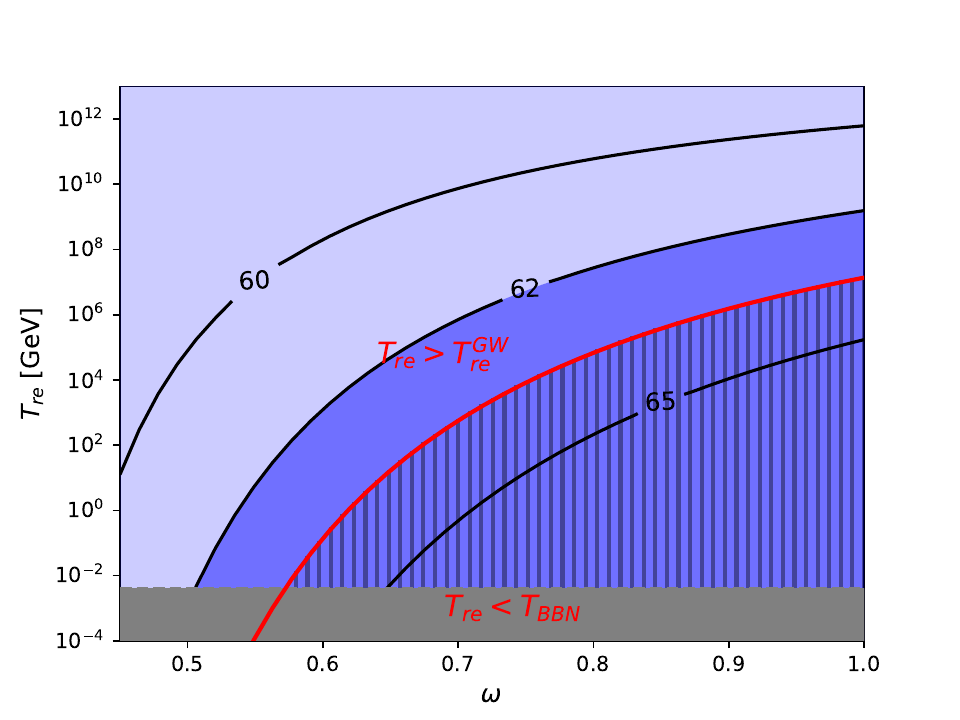}}
	\caption{\label{Tre_plots}
		Reheating temperature $T_{\mathrm{re}}$ as a function of the effective equation-of-state parameter $\omega_{\mathrm{re}}$ 
		for two representative Tsallis holographic parameter sets, 
		with $(\alpha,\beta,\delta)=(0.01,-11,2.5)$ for (a) and $(0.03,-10,5)$ for (b), and fixed $\lambda=5.0$.
		The lower horizontal line denotes the BBN limit ($T_{\mathrm{re}} = 4\,\mathrm{MeV}$),  
		while the upper line shows the PGW-induced lower bound $T_{\mathrm{re}}^{\mathrm{GW}}$ derived in Eq.~\eqref{eq:TreGW_standard}, relevant when $\omega_{\mathrm{re}}>1/3$.  
		Allowed reheating histories correspond to temperatures that lie above both lines and are consistent with Eq.~\eqref{eq:Tre_closed}.}
\end{figure}

Figure~\ref{Tre_plots}(a) shows the reheating temperature $T_{\mathrm{re}}$ as a function of the effective equation-of-state parameter $\omega_{\mathrm{re}}$ for the parameter set $(\alpha,\beta,\delta,\lambda)=(0.01,-11,2.5,5)$. % and for $N_k=58,60,$ and $66$. 
The dark-blue color area indicates that the results of the model for the chosen sets of parameters and for the value of inflationary e-folds remain consistent with the $68\%$CL of ACT DR6 data, and the light-blue color area shows consistency with the $95\%$CL. 
The red-sold curve is the constraint temperature $T_{\mathrm{re}}^{\mathrm{GW}}$. One can realized that this new constraint become efficient for $\omega_{re} > 0.57$, and for lower $\omega_{\rm re}$, the BBN temperature becomes the efficient bound. The black-sold lines show the contour line of inflationary e-folds, and the shaded area corresponds to the reheating temperature below $T_{\mathrm{re}}^{\mathrm{GW}}$. As expected from Eq.~\eqref{eq:Tre_closed}, increasing $N_k$ shifts the reheating curves downward. Over the full plotted range of $\omega_{\mathrm{re}}$, it is realized that for $N_k \lesssim 62$, the resulting reheating temperature are above the $T_{\mathrm{re}}^{\mathrm{GW}}$ constraint line, and the condition \eqref{eq:TreGW_standard} is satisfied. However, by higher values of inflationary e-folds, $N_k > 62$, the reheating temperature falls below the PGW-induced bound $T_{\mathrm{re}}^{\mathrm{GW}}$ and the condition \eqref{eq:TreGW_standard} is violated. Additionally, the reheating temperature drops below the BBN limit for larger values of $ N_k$. For $N_k = 66$, it falls below the BBN limit for $\omega_{\rm re} < 0.72$, and violates both constraints simultaneously. Hence, for this Tsallis holographic parameter set, values $N_k > 62$ are disfavored. \\
%For $N_k=58$ and $N_k=60$, the reheating temperature remains above the BBN limit $T_{\mathrm{BBN}}\simeq4~\mathrm{MeV}$ over the full plotted range of $\omega_{\mathrm{re}}$. In contrast, for $N_k=66$ the curve falls below the PGW-induced bound $T_{\mathrm{re}}^{\mathrm{GW}}$, \textcolor{blue}{which in this region also lies above the BBN threshold and therefore violates both constraints simultaneously},for $\omega_{\mathrm{re}}\gtrsim0.6$. Hence, for this Tsallis holographic parameter set, values $N_k\gtrsim60$ are disfavored.\\
Figure~\ref{Tre_plots}(b) displays the same analysis for the second parameter set
$(\alpha,\beta,\delta,\lambda)=(0.03,-10,5,5)$, where the black-lines show the contour lines related to different values of inflationary e-folds. The red-solid line displays the $T_{\mathrm{re}}^{\mathrm{GW}}$ constraint, which is efficient for stiff reheating with $\omega_{\rm re} \gtrsim 0.5$. In this case, the reheating temperature remains safely above both the BBN and PGW bounds for $N_k \lesssim 63$. While for $N_k < 62$, the resulting temperature perfectly satisfies both constraints, the model prediction of the inflationary phase remains in the $95\%$CL of ACT DR6. Therefore, for this case, there is a stronger constraint on the inflationary e-folds to satisfy both ACT DR6 for the inflationary phase and reheating temperature constraints simultaneously. There are similar conclusion compared to the previous case, so that by increasing the inflationary e-folds $N_k$, the reheating temperature falls below PGW-induced bound $T_{\mathrm{re}}^{\mathrm{GW}}$, and for some specific value of the plotted range of $\omega_{\rm re}$ it even drops below the BBN temperature; signaling an inconsistent reheating history. 
%while for $N_k=65$ the curve clearly enters the region excluded by $T_{\mathrm{re}}^{\mathrm{GW}}$. \textcolor{blue}{Moreover, for $\omega_{\mathrm{re}}\gtrsim0.6$, the reheating temperature drops below the BBN limit, signaling an inconsistent reheating history.}
%Hence, the allowed inflationary duration is restricted to $N_k\lesssim62$.
%%%%%%%%%%%%%%%%%%%%%%%%%%%%%%%%%%%%
%%%%%%%%%%%%%%%%%%%%%%%%%%%%%%%%%%%%

\begin{figure}[htbp]
	\centering 
	\subfigure[\label{GW01}]{\includegraphics[width=0.4\linewidth]{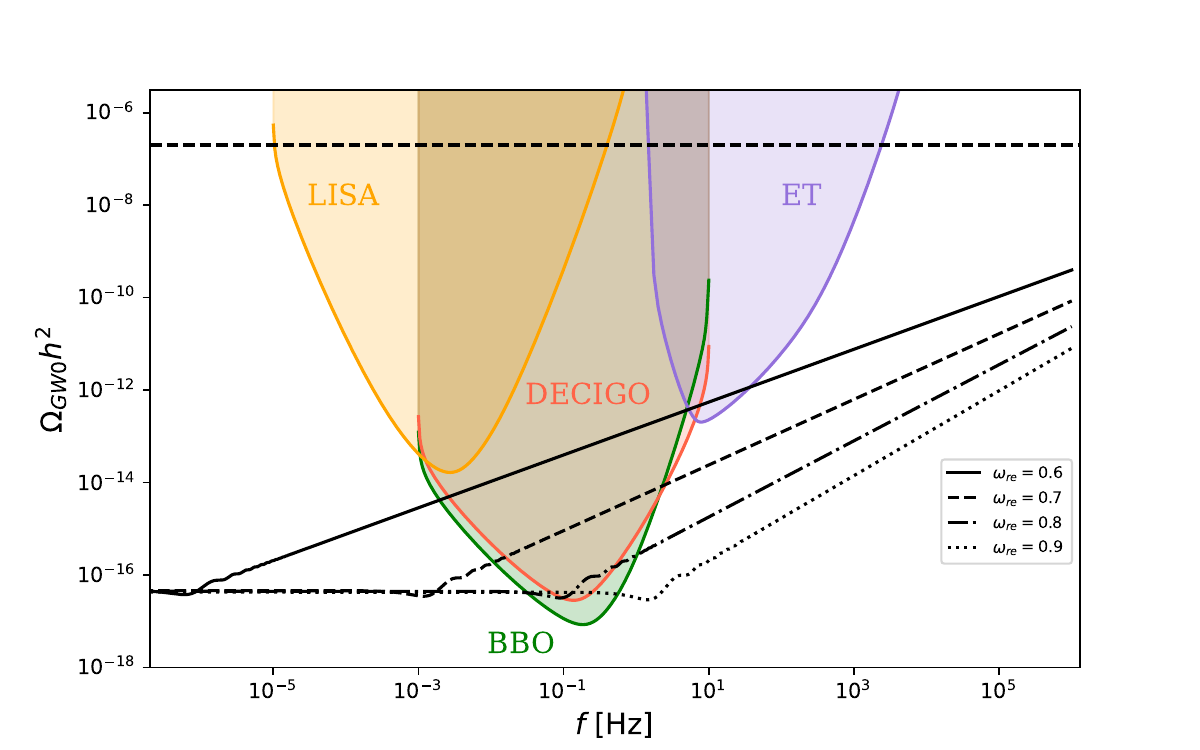}}
	\subfigure[\label{GW02}]{\includegraphics[width=0.4\linewidth]{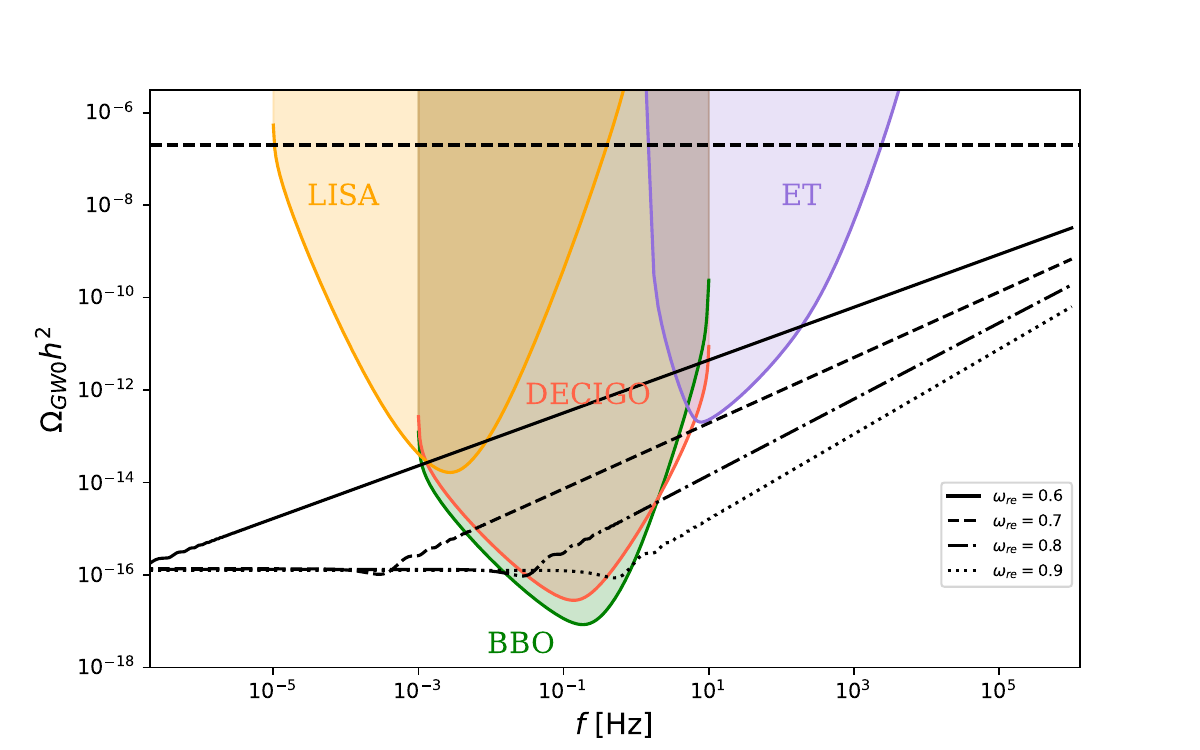}}
	\caption{\label{GW_plots}
		Present-day primordial GW spectrum $\Omega_{\mathrm{GW}}^{(0)}(k)$
		for the same parameter sets as in Fig.~\ref{Tre_plots}. 
		For $\omega_{\mathrm{re}}>1/3$, the high-frequency slope steepens according to the power-law index
		$(6\omega_{\mathrm{re}} - 2)/(1 + 3\omega_{\mathrm{re}})$ in Eq.~\eqref{eq:PGW_spectrum},  
		and the turnover scale is set by $k_{\mathrm{re}}$ through the reheating relation in Eq.~\eqref{eq:Tre_closed}.  
		Sensitivity curves of future detectors are included for comparison.}
\end{figure}

The corresponding present-day PGWs energy spectra are shown in Fig.~\ref{GW_plots} versus the frequency and for different values of the effective reheating equation of state, $\omega_{\mathrm{re}}=0.6,0.7,0.8,$ and $0.9$ indicated by solid, dashed, dot-dashed, and dotted black lines, respectively. At low frequencies, the spectra are nearly scale invariant, while at high frequencies a pronounced blue tilt develops in agreement with Eq.~\eqref{eq:PGW_spectrum}. 
For the first parameter set [Fig.~\ref{GW_plots}(a)], spectra with $\omega_{\mathrm{re}}\ge 0.6$ and $0.7$ enter the sensitivity band of DECIGO and BBO, while spectra with $\omega_{\mathrm{re}}\ge 0.8$ and $0.9$ cross the sensitivity bands of BBO, and they tough the sensitivity band of DECIGO. 
For the second set [Fig.~\ref{GW_plots}(b)], the resulting spectra in general cross the observable bands for a wider range, so that for all plotted values of $\omega_{\rm re}$, they cross the BBO and DECIGO observable range. Additionally, for $\omega_{\rm re}$, the resulting energy spectrum reaches the sensitivity bands of ET, and it also touches the observable range of LISA as well. \\  %All spectra satisfy $\Delta N_{\mathrm{eff}}\le0.17$.
%%%%%%%%%%%%%%%%%%%%%%%%%%%%%%%%%%%%
%%%%%%%%%%%%%%%%%%%%%%%%%%%%%%%%%%%%
Taken together, inflation, reheating, and PGWs significantly sharpen the viable region of the Tsallis holographic inflationary parameter space. Only those combinations of $(\alpha,\beta,\delta,\lambda)$ that yield sufficiently large $T_{\mathrm{re}}$ to satisfy both the BBN constraint and the PGWs-induced constraint $T_{\mathrm{re}}<T_{\mathrm{re}}^{\mathrm{GW}}$. 
In particular, while the first parameter set requires $N_k \lesssim 62$, the second set is restricted to $N_k \lesssim 63$. The inflationary e-folds $N_k \simeq65$ yields $T_{\mathrm{re}}<T_{\mathrm{re}}^{\mathrm{GW}}$ for $\omega_{\mathrm{re}} \gtrsim 0.6$ that violate the condition.

%%%%%%%%%%%%%%%%%%%%%%%%%%%%%%%%%%%%%%%%%%%%%%%
%%%%%%%%%%%%%%%%%%%%%%%%%%%%%%%%%%%%%%%%%%%%%%%
%%%%%%%%%%%%%%%%%%%%%%%%%%%%%%%%%%%%%%%%%%%%%%%
%%%%%%%%%%%%%%%%%%%%%%%%%%%%%%%%%%%%%%%%%%%%%%%
% ============ Sec.VI =======================================
%%%%%%%%%%%%%%%%%%%%%%%%%%%%%%%%%%%%%%%%%%%%%%%
%%%%%%%%%%%%%%%%%%%%%%%%%%%%%%%%%%%%%%%%%%%%%%%
%%%%%%%%%%%%%%%%%%%%%%%%%%%%%%%%%%%%%%%%%%%%%%%
%%%%%%%%%%%%%%%%%%%%%%%%%%%%%%%%%%%%%%%%%%%%%%%
\clearpage
\section{Scalar field, potential, and swampland criteria}\label{reconstructing_potential}

In the previous section, the THDE was used to construct an effective inflationary potential expressed in terms of the Hubble parameter. This provides a convenient starting point to reconstruct the scalar-field dynamics %in a model-independent way and to assess the compatibility of the scenario with quantum-gravity, motivated swampland criteria. 
Compatibility of the model with ACT DR6 was examined and we found the acceptable ranges for the free parameters of the model. In this section we use that result to find the scalar field and reconstruct the potential in the slow-roll regime, and additionally explore the validity of the swampland criteria for the model, which are a set of conjecture consistency conditions originated from string theory \cite{Obied:2018sgi,Garg:2018reu,Ooguri:2018wrx}. \\
Using Eq.~\eqref{Fried2_field_sr}, the kinetic term of the inflaton can be written in terms of the Hubble evolution as
\begin{equation}\label{phi_dot}
	\dot{\phi}^{2} = -\,\frac{2 M_{p}^{2}}{1+\lambda}\,\dot{H}.
\end{equation}
The total field excursion then follows from
\begin{equation}
	\Delta\phi = |\phi_\star - \phi_e| = \int_{N_e}^{N_\star} \left| \frac{d\phi}{dN} \right| dN = \int_{N_e}^{N_\star} \frac{|\dot{\phi}|}{H}  dN .
\end{equation}
where $\Delta\phi$ quantifies the total field displacement between horizon exit of the pivot scale and the end of inflation. This quantity plays a central role in the distance conjecture. In this section, $N$ denotes the number of $e$-folds from horizon exit ($N = N_\star$) to the end of inflation ($N = N_e$). 

\begin{figure}[htbp]
	\centering
	\includegraphics[width = 0.4\linewidth]{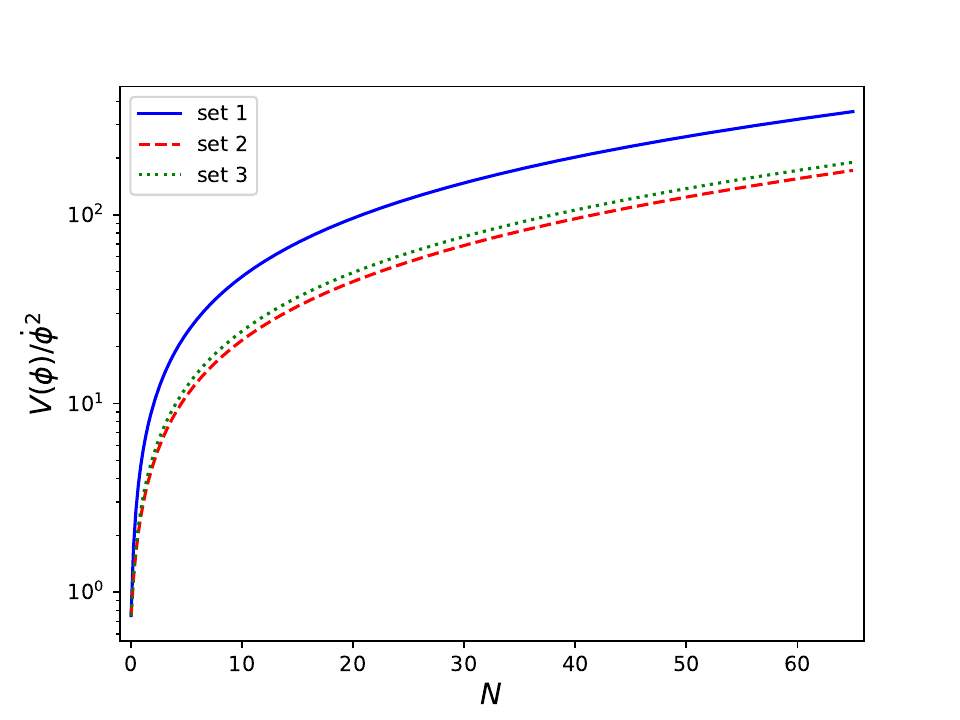}
	\caption{
		Ratio $V/\dot\phi^{2}$ for three Tsallis parameter sets:
		set1 = $(0.01,-11,2.5)$,
		set2 = $(0.03,-10,5)$,
		set3 = $(0.02,-8,4.0)$.
		The matter--geometry coupling is fixed to $\lambda = 10^{3}$.
		In all cases $V\gg\dot\phi^{2}$ across the observable inflationary window, ensuring the validity of the slow-roll approximation.}
	\label{pot_kinetic_ratio}
\end{figure}

From \eqref{phi_dot}, one can extract the kinetic energy density of the scalar field, and the potential part can be read from \eqref{fried_hde}. Fig.~\ref{pot_kinetic_ratio} illustrates the ratio of the potential energy density over the kinetic energy during the inflationary phase for three sets of $(\alpha, \beta, \delta)$ where the matter--geometry coupling fixed at $\lambda=10^{3}$. Although the parameter sets differ in their Tsallis parameters, in all cases, the slow-roll approximation $V \gg \dot{\phi}^{2}$ is satisfied, so that it remains of the order of $\mathcal{O}(10^2)$; confirming the robustness of slow roll. It should be note that this result does not depends on the high value of $\lambda$ parameter and there is the same result as one take $\lambda \sim \mathcal{O}(1)$. In the following lines, we explain that a large matter-geometry coupling parameter $\lambda$ is required to satisfy the swampland criteria.  
%In the following, we will therefore treat $\Delta\phi$ and the normalized slope $V'/V$ as the primary diagnostics of swampland compatibility. The slow-roll approximation requires $V \gg \dot{\phi}^{2}$, and Fig.~\ref{pot_kinetic_ratio} illustrates this hierarchy for three representative choices of $(\alpha,\beta,\delta)$ with the matter--geometry coupling fixed at $\lambda=10^{3}$. Although the parameter sets differ in their Tsallis parameters, in all cases $V/\dot{\phi}^{2}$ remains between one and two orders of magnitude throughout the observable window, confirming the robustness of slow roll. Since $\lambda$ is kept fixed in this figure, the variations among the curves reflect only the changes in $(\alpha,\beta,\delta)$. The choice $\lambda = 10^{3}$ is made solely to enhance the visual separation between the kinetic and potential contributions and is not used in the quantitative swampland analysis presented below.
%%%%%%%%%%%%%%%%%%%%%%%%%%%%%%%%%%%%
%%%%%%%%%%%%%%%%%%%%%%%%%%%%%%%%%%%%

%%%%%%%%%%%%%%%%%%%%%%%%%%%%%%%%%%%%%%%%%%%%%%%
%%%%%%%%%%%%%%%%%%%%%%%%%%%%%%%%%%%%%%%%%%%%%%%
%%%%%%%%%%%%%%%%%%%%%%%%%%%%%%%%%%%%%%%%%%%%%%%
\subsection{Swampland criteria}

In addition to observational consistency, a viable inflationary model must satisfy theoretical constraints inspired by quantum gravity. The swampland conjectures \cite{Obied:2018sgi,Garg:2018reu,Ooguri:2018wrx} provide criteria intended to distinguish effective field theories that can arise from a consistent ultraviolet completion. For single-field inflation, the relevant conditions can be summarized as follows,
\begin{enumerate}
    \item The distance conjecture requires a sub-Planckian field excursion,
        \begin{equation}
        	\frac{\Delta\phi}{M_{p}} \leq c_{1},
        \end{equation}
        where $c_{1}$ is a constant of order unity.

    \item The de Sitter (gradient) conjecture demands a sufficiently steep potential,
        \begin{equation}
        	\frac{1}{M_{p}}\frac{|V'|}{V} \geq c_{2},
        \end{equation}
        with $c_{2} \sim \mathcal{O}(1)$. In its refined form, at least one of the following must hold:
        \begin{equation}
        	\frac{1}{M_p}\frac{|V'|}{V} \ge c_2, 
        	\qquad\text{or}\qquad 
        	\frac{1}{M_p^{2}}\frac{V''}{V} \le -\,c_{2}' ,
        \end{equation}
        where the second inequality states that the second derivative of the potential with respect to the scalar field must be sufficiently negative (tachyonic), as emphasized in Ref.~\cite{Obied:2018sgi}. 
\end{enumerate}

\begin{figure}[htbp]
	\centering
	\includegraphics[width = 0.4\linewidth]{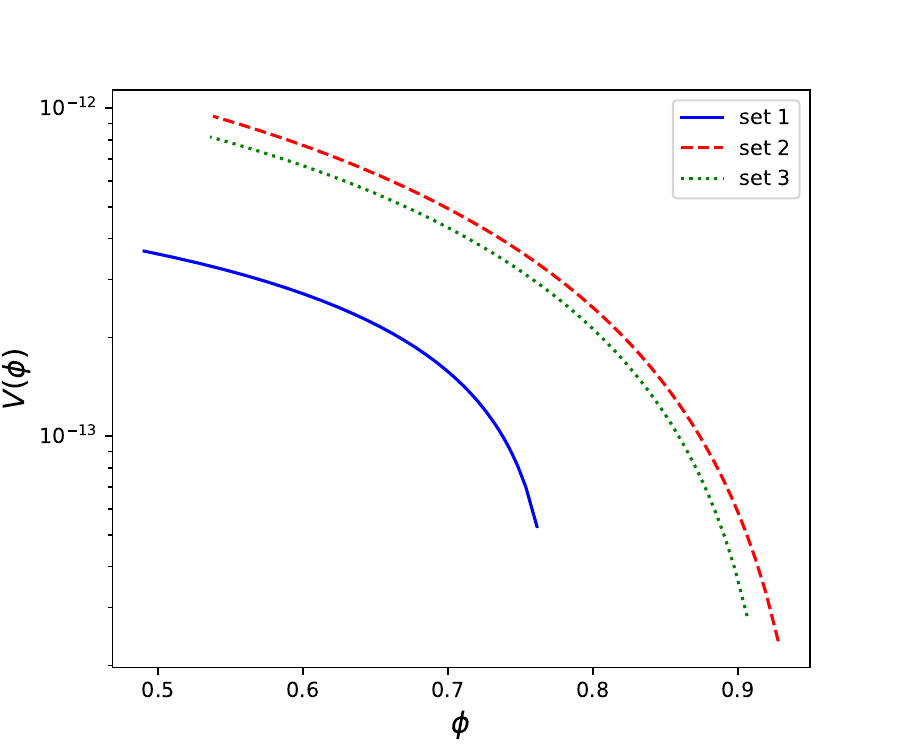}
	\caption{
		Reconstructed inflaton potential $V(\phi)$ for the same parameter sets as in Fig.~\ref{pot_kinetic_ratio}, with $\lambda = 10^{3}$ fixed.
		The curves exhibit a plateau at large field values and a smooth descent toward the end of inflation.
		The displayed field interval corresponds to the observable inflationary window and does not cover the entire excursion $\Delta\phi$.}
	\label{pot_phi}
\end{figure}

\begin{figure}[htbp]
	\centering
	\includegraphics[width = 0.4\linewidth]{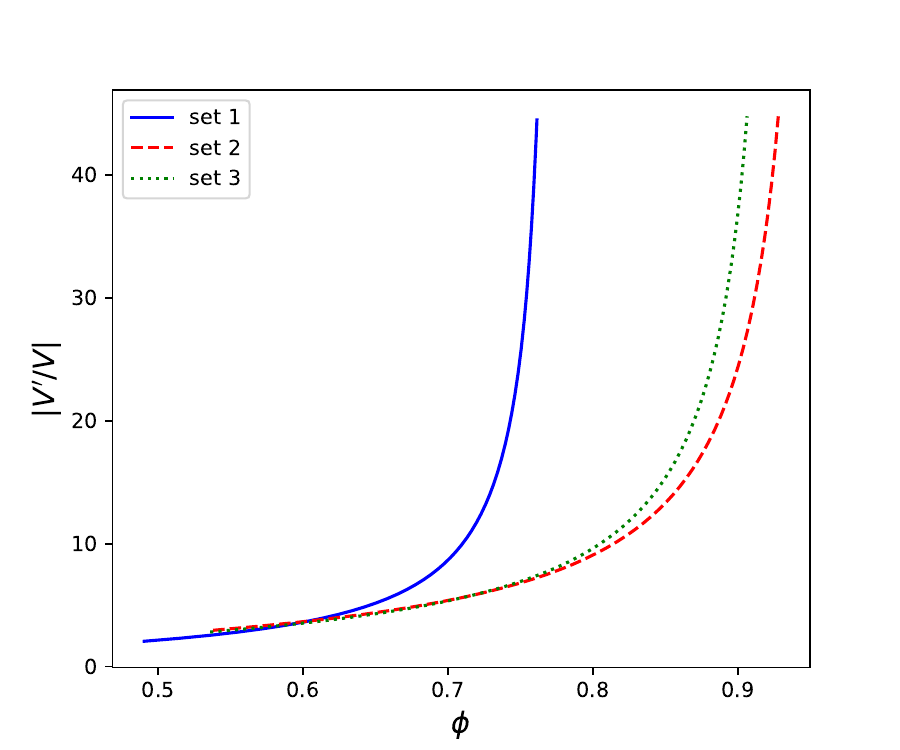}
	\caption{
		Evolution of the normalized potential slope $V'/V$ during inflation for the same parameter sets used in Figs.~\ref{pot_kinetic_ratio}, with $\lambda=10^{3}$ fixed.
		The increase in $V'/V$ toward the end of inflation signals the departure from the plateau regime and the approach to the end of slow roll.}
	\label{sc_second}
\end{figure}

In practice, we focus on the distance during the inflationary time and gradient conditions evaluated at the pivot-scale horizon exit, where the inflationary observables are defined. Integrating $\dot{\phi}$ gives the scalar-field trajectory, which combined with $V(H)$ yields a parametric reconstruction of $V(\phi)$. Figure~\ref{pot_phi} shows the resulting potential for the same three parameter sets. A plateau-like region appears at small field values, followed by a monotonic increase as the field approaches the end of inflation. %Since $\lambda=10^3$ is held fixed, 
This figure provides a qualitative illustration of the potential shape and its on the parameters of the model. %dependence on $\lambda$ will be analyzed using Table~\ref{table_sc}. The field range displayed in Fig.~\ref{pot_phi} corresponds to the observable $e$-fold interval rather than the full field excursion $\Delta\phi$ and is chosen to cover the scales relevant for CMB observations. 
From the figure, one finds that the field range during the inflationary phase remains smaller than unity which is in consistency with the first swampland criterion. Depending on the free parameters, the potential could get a higher amplitude and the field distance can be affected as well. It should be noted that for small value of the matter--geometry coupling $\lambda$, the field distance during the inflationary time is always larger than one, and to satisfy the distance conjecture it is required that the coupling parameter $\lambda$ be at lest of the order of $\mathcal{O}(10^2)$. This conclusion can be seen in Table~\ref{table_sc}. \\ 
%%%%%%%%%%%%%%%%%%%%%%%%%%%%%%%%%%%%
%%%%%%%%%%%%%%%%%%%%%%%%%%%%%%%%%%%%
%A further quantity of interest is the normalized slope of the potential, $V'/V$, which is directly related to the slow-roll parameter $\epsilon_V$ via $\epsilon_V \simeq \tfrac{M_p^2}{2}(V'/V)^2$.
Figure~\ref{sc_second} displays the evolution of $V'/V$ during inflation for the same Tsallis parameter sets and $\lambda=10^3$. In all cases the slope increases as inflation progresses, reflecting the departure from the quasi-de Sitter plateau near the end of inflation. This is the same result for all chosen sets of parameter. %Again, since $\lambda$ is fixed, this figure illustrates only the qualitative behavior; 
The quantitative impact of varying $\lambda$ will be extracted from the numerical values shown in Table~\ref{table_sc}. In particular, the values of $V'/V$ at the pivot scale will be used to test the gradient form of the swampland de Sitter conjecture.
%%%%%%%%%%%%%%%%%%%%%%%%%%%%%%%%%%%%
%%%%%%%%%%%%%%%%%%%%%%%%%%%%%%%%%%%%
While Figs.~\ref{pot_phi} and \ref{sc_second} illustrate qualitatively how $V(\phi)$ and $V'/V$ behave during inflation, the quantitative dependence of the field excursion and slope on the matter--geometry coupling $\lambda$ is encoded in Table~\ref{table_sc}. The field distance is measure during the whole inflationary times and the gradient of the potential is evaluated at the pivot-scale horizon exit, corresponding to $N=65$. This choice lies within the standard range of $e$-folds compatible with CMB observations and reheating constraints. From the Table~\ref{table_sc}, it is realized that for $\lambda=8$, the entries satisfy
\begin{equation}
	\frac{\Delta\phi}{M_p} \simeq 3.5{-}4.0,
	\qquad 
	\frac{1}{M_{p}}\frac{|V'|}{V} \simeq 0.26{-}0.32,
\end{equation}
indicating that neither the distance conjecture nor the gradient conjecture is satisfied. 
In contrast, for $\lambda=200$ the table shows
\begin{equation}
	\frac{\Delta\phi}{M_p} \simeq 0.75{-}0.85,
	\qquad 
	\frac{1}{M_{p}}\frac{|V'|}{V} \simeq 1.20{-}1.48,
\end{equation}
so that both swampland conditions are satisfied simultaneously. The dependence on $(\alpha,\beta,\delta)$ is comparatively mild: varying these parameters changes the detailed shape of $V(\phi)$ but does not alter the order of magnitude of $\Delta\phi$ and $V'/V$. By contrast, the transition from violation to satisfaction of the conjectures is governed primarily by the strength of the matter--geometry coupling.  Taken together, these results indicate that the threshold value required for compatibility with swampland constraints is
$\lambda \gtrsim \mathcal{O}(10^{2})$.
We emphasize that this swampland-compatible range of $\lambda$ is also consistent with the observational bounds on the scalar spectral index and tensor-to-scalar ratio obtained in the previous section, so that the Tsallis holographic inflation model in $f(R,T)$ gravity remains simultaneously viable from both phenomenological and theoretical perspectives.
%%%%%%%%%%%%%%%%%%%%%%%%%%%%%%%%%%%%
\begin{table}[t]
	\centering
	\setlength{\tabcolsep}{10pt}
	\renewcommand{\arraystretch}{1.2}
	\caption{Field excursion $\Delta\phi$ and normalized potential slope $V'/V$ for several $(\lambda,\alpha,\beta,\delta)$ combinations in the Tsallis holographic inflation model at $N = 65$ and $M_{p}=1$.}
	\label{table_sc}
	\begin{tabular}{cccccc}
		\hline
		$\lambda$ & $\alpha$ & $\beta$ & $\delta$ & $\Delta\phi$ & $V'/V$ \\
		\hline
		8.0 & 0.01 & -12.0 & 3.5 & 3.671 & 0.287 \\
		8.0 & 0.01 & -12.0 & 4.5 & 3.990 & 0.315 \\
		8.0 & 0.01 & -10.0 & 3.5 & 3.660 & 0.284 \\
		8.0 & 0.01 & -10.0 & 4.5 & 3.979 & 0.313 \\
		8.0 & 0.01 & -8.0  & 3.5 & 3.644 & 0.281 \\
		8.0 & 0.01 & -8.0  & 4.5 & 3.964 & 0.309 \\
		8.0 & 0.02 & -12.0 & 3.5 & 3.616 & 0.275 \\
		8.0 & 0.02 & -12.0 & 4.5 & 3.939 & 0.304 \\
		8.0 & 0.02 & -10.0 & 3.5 & 3.594 & 0.270 \\
		8.0 & 0.02 & -10.0 & 4.5 & 3.919 & 0.299 \\
		8.0 & 0.02 & -8.0  & 3.5 & 3.562 & 0.263 \\
		8.0 & 0.02 & -8.0  & 4.5 & 3.889 & 0.293 \\
		8.0 & 0.03 & -12.0 & 3.5 & 3.562 & 0.263 \\
		8.0 & 0.03 & -12.0 & 4.5 & 3.889 & 0.293 \\
		8.0 & 0.03 & -10.0 & 3.5 & 3.529 & 0.256 \\
		8.0 & 0.03 & -10.0 & 4.5 & 3.859 & 0.286 \\
		8.0 & 0.03 & -8.0  & 3.5 & 3.481 & 0.245 \\
		8.0 & 0.03 & -8.0  & 4.5 & 3.814 & 0.276 \\
		200.0 & 0.01 & -12.0 & 3.5 & 0.777 & 1.355 \\
		200.0 & 0.01 & -12.0 & 4.5 & 0.844 & 1.488 \\
		200.0 & 0.01 & -10.0 & 3.5 & 0.775 & 1.344 \\
		200.0 & 0.01 & -10.0 & 4.5 & 0.842 & 1.477 \\
		200.0 & 0.01 & -8.0  & 3.5 & 0.771 & 1.326 \\
		200.0 & 0.01 & -8.0  & 4.5 & 0.839 & 1.461 \\
		200.0 & 0.02 & -12.0 & 3.5 & 0.765 & 1.298 \\
		200.0 & 0.02 & -12.0 & 4.5 & 0.833 & 1.435 \\
		200.0 & 0.02 & -10.0 & 3.5 & 0.761 & 1.275 \\
		200.0 & 0.02 & -10.0 & 4.5 & 0.829 & 1.414 \\
		200.0 & 0.02 & -8.0  & 3.5 & 0.754 & 1.241 \\
		200.0 & 0.02 & -8.0  & 4.5 & 0.823 & 1.383 \\
		200.0 & 0.03 & -12.0 & 3.5 & 0.754 & 1.241 \\
		200.0 & 0.03 & -12.0 & 4.5 & 0.823 & 1.383 \\
		200.0 & 0.03 & -10.0 & 3.5 & 0.747 & 1.208 \\
		200.0 & 0.03 & -10.0 & 4.5 & 0.816 & 1.352 \\
		200.0 & 0.03 & -8.0  & 3.5 & 0.737 & 1.158 \\
		200.0 & 0.03 & -8.0  & 4.5 & 0.807 & 1.306 \\
		\hline
	\end{tabular}
\end{table}
%%%%%%%%%%%%%%%%%%%%%%%%%%%%%%%%%%%%
These results demonstrate that increasing $\lambda$ suppresses the field excursion while enhancing the gradient of the potential. In the limit of large $\lambda$, and in particular for $\lambda \gtrsim \mathcal{O}(10^{2})$, the model naturally satisfies both the distance and gradient conjectures, thereby lying within the swampland-compatible region of parameter space.

%%%%%%%%%%%%%%%%%%%%%%%%%%%%%%%%%%%%%%%%%%%%%%%
%%%%%%%%%%%%%%%%%%%%%%%%%%%%%%%%%%%%%%%%%%%%%%%
%%%%%%%%%%%%%%%%%%%%%%%%%%%%%%%%%%%%%%%%%%%%%%%
%%%%%%%%%%%%%%%%%%%%%%%%%%%%%%%%%%%%%%%%%%%%%%%
% ============ Sec.VII =======================================
%%%%%%%%%%%%%%%%%%%%%%%%%%%%%%%%%%%%%%%%%%%%%%%
%%%%%%%%%%%%%%%%%%%%%%%%%%%%%%%%%%%%%%%%%%%%%%%
%%%%%%%%%%%%%%%%%%%%%%%%%%%%%%%%%%%%%%%%%%%%%%%
%%%%%%%%%%%%%%%%%%%%%%%%%%%%%%%%%%%%%%%%%%%%%%%
\clearpage
\section{Conclusions and Outlook}\label{conclusion}

In this work, we have demonstrated that THDE embedded in $f(R,T)$ gravity provides a predictive, theoretically motivated, and observationally consistent framework for early-universe inflation. 
By combining the non-extensive entropy relation with the GO infrared cutoff, the resulting potential acquires an explicitly holographic form $V(\phi)$ determined by the background dynamics. When supplemented with the slow-roll equations of $f(R,T)$ gravity, this structure yields a tightly predictive scenario in which the inflationary observables are governed by the parameters $(\alpha,\beta,\delta,\lambda)$ and the number of $e$-folds.

A detailed comparison with ACT DR6 (P-ACT-LB) likelihoods shows that the model accommodates a broad region of parameter space consistent with current CMB constraints. Viable solutions favor moderately large non-extensivity, $\delta \gtrsim 2$, and negative GO coefficient $\beta$. Representative points in Table~\ref{table_rns} lead to
$0.9687 \lesssim n_s \lesssim 0.973$, and  
$0.0059 \lesssim r \lesssim 0.024 $,
which lie comfortably within the $68\%$ CMB confidence region. These results place the THDE--$f(R,T)$ construction securely inside the observationally preferred region across distinct parameter choices. 

Reconstructing the scalar field and its potential determined that, throughout the observable window, the ratio $V/\dot{\phi}^{2}$ remains of order $10^{2}$. The reconstructed potential $V(\phi)$ in Fig.~\ref{pot_phi} exhibits a controlled plateau that transitions smoothly toward the end of inflation. 
%The scalar-sector reconstruction further confirms the internal consistency of the framework. Throughout the observable window, the ratio $V/\dot{\phi}^{2}$ remains of order $10^{2}$, while the reconstructed potential $V(\phi)$ in Fig.~\ref{pot_phi} exhibits a controlled plateau that transitions smoothly toward the end of inflation. 
%The potential gradient normalized by the potential value, quantified by $|V'|/(V M_{p})$, increases monotonically as the inflaton rolls, indicating a progressively steepening holographic potential. Both the field excursion and $|V'|/(V M_{p})$ are predominantly determined by the matter--geometry coupling $\lambda$. 
Examining the swampland criteria determined that both the distance and de-Sitter conjectures are violated for small values of the matter--geometry coupling parameter $\lambda$; however, as the parameter gets of the order of $10^2$ (or larger), both conjectures will be satisfied. As shown in Table~\ref{table_sc}, increasing $\lambda$ suppresses $\Delta\phi/M_p$ and enhances $|V'|/(V M_{p})$, allowing the distance and gradient swampland conjectures to be satisfied simultaneously for $\lambda \gtrsim \mathcal{O}(10^{2})$. 

Since the universe is in an extremely cold state and almost empyt of standard particle, a phase of reheating is required to heat up the universe, fill it with particle, and recover the radiation-dominant phase. This phase is oftenly parametrized by three parameters as the reheating e-folds $N_{\rm re}$, reheating temperture $T_{\rm re}$, and reheating effective equation of state $\omega_{\rm re}$. 
%\textcolor{red}{The post-inflationary evolution is tightly constrained by the reheating phase and the associated primordial gravitational-wave (PGW) background.} 
Successful BBN imposes the lower bound $T_{\mathrm{re}}\gtrsim4\,\mathrm{MeV}$, while the P-ACT-LB limit on excess radiation, $\Delta N_{\mathrm{eff}}\le0.17$, provides a complementary constraint that becomes particularly restrictive for stiff reheating with $\omega_{\mathrm{re}}>0.5$.  
%Using the model-independent relations between $T_{\mathrm{re}}$, $N_k$, and $\omega_{\mathrm{re}}$, we showed that reheating significantly narrows the otherwise viable inflationary parameter space.\\
For the Tsallis holographic parameter sets $(\alpha,\beta,\delta,\lambda)=(0.01,-11,2.5,5)$ and $(0.03,-10,5,5)$, the reheating temperature was studied and it was showed that it remains in consistency with both the BBN bound and the PGW-induced limit $T_{\mathrm{re}}^{\mathrm{GW}}$ only for $N_k \lesssim 62$ and $N_k \lesssim 63$, respectively. For larger inflationary durations, in particular $N_k\simeq65$--$66$, the reheating curves fall below $T_{\mathrm{re}}^{\mathrm{GW}}$ for $\omega_{\mathrm{re}}\gtrsim0.6$, thereby simultaneously violating the BBN and PGW constraints. This demonstrates that the reheating phase imposes a sharp upper bound on the inflationary duration once PGW consistency is enforced.\\
The corresponding present-day PGW spectra exhibit a characteristic blue tilt for $\omega_{\mathrm{re}}>1/3$, with a turnover scale governed by the reheating temperature. For $\omega_{\mathrm{re}} \ge 0.6$, the predicted signals enter the projected sensitivity ranges of upcoming space-based gravitational-wave detectors, including DECIGO and LISA, while remaining fully consistent with the bound $\Delta N_{\mathrm{eff}}\le0.17$. 
Taken together, reheating and PGW constraints act as a powerful dynamical filter that significantly sharpens the predictive power of the THDE--$f(R,T)$ framework.

%In summary, the THDE--$f(R,T)$ scenario provides a coherent, theoretically grounded, and observationally viable model of early-universe dynamics, with predictions that naturally link \textcolor{red}{ultraviolet consistency}, reheating physics, and the high-frequency PGW background. Future observations, including CMB-S4, LiteBIRD, next-generation interferometers, and pulsar-timing arrays, will directly probe the parameter space identified here and offer concrete opportunities to test or falsify this framework. 
%Potential extensions include studies of primordial non-Gaussianity, multifield or warm-inflation generalizations, non-canonical kinetic sectors, and non-standard reheating histories. These directions may further illuminate the role of holographic entropy and matter--geometry coupling in the quantum structure of the early universe.

%%%%%%%%%%%%%%%%%%%%%%%%%%%%%%%%%%%%%%%%%%%%%%%
\appendix	
%%%%%%%%%%%%%%%%%%%%%%%%%%%%%%%%%%%%%%%%%%%%%%%

%%%%%%%%%%%%%%%%%%%%%%%%%%%%%%%%%%%%%%%%%%%%%%%
\end{document}